\documentclass[aps,prl,twocolumn,notitlepage,showpacs,superscriptaddress,am]{revtex4-1}%
\usepackage{graphicx}
\usepackage{amsmath,amsfonts,amssymb}
\usepackage{color}
\newcommand{\nairo}{Na$_2$IrO$_3$}
\newcommand{\liiro}{Li$_2$IrO$_3$}

\newcommand{\hliiro}{H$_3$LiIr$_2$O$_6$}

\begin{document}

\title{Role of Hydrogen  in the spin-orbital-entangled
quantum liquid candidate {\hliiro}}
\date{\today}
\begin{abstract}
 Motivated by  recent reports of  
 {\hliiro} as a spin-orbital-entangled
quantum liquid,
we investigate via a combination of density functional theory and 
non-perturbative exact diagonalization the microscopic nature of 
its magnetic interactions. 
We find that  while the interlayer O-H-O bond geometry
strongly affects the local magnetic couplings,  these  bonds
are likely to remain symmetrical due to large zero-point
fluctuations of the H positions. In this case, 
 the estimated magnetic model 
lies close to the classical tricritical point between ferromagnetic,
zigzag and incommensurate spiral orders, what may contribute to the lack of
magnetic ordering.
 However, we also find that
 substitution of H by D (deuterium) as well as 
 disorder-induced inhomogeneities 
destabilize the O-H/D-O
bonds  modifying strongly the local magnetic couplings. These
results suggest that  the magnetic response in {\hliiro} is
likely sensitive to both the stoichiometry and microstructure of
the samples and emphasize the importance of a careful treatment of hydrogen for similar systems.  \end{abstract}

\author{Ying Li} \altaffiliation{Corresponding author: yingli@itp.uni-frankfurt.de}
\affiliation{Institut f\"ur Theoretische Physik, Goethe-Universit\"at Frankfurt,
Max-von-Laue-Strasse 1, 60438 Frankfurt am Main, Germany}
\author{Stephen M. Winter}
\affiliation{Institut f\"ur Theoretische Physik, Goethe-Universit\"at Frankfurt,
Max-von-Laue-Strasse 1, 60438 Frankfurt am Main, Germany}
\author{Roser Valent{\'\i}}
\affiliation{Institut f\"ur Theoretische Physik, Goethe-Universit\"at Frankfurt,
Max-von-Laue-Strasse 1, 60438 Frankfurt am Main, Germany}
\maketitle
Since the proposal by A. Kitaev   of
a $Z_2$ spin liquid groundstate
in a honeycomb lattice with bond-dependent Ising-like nearest neighbor
interactions~\cite{Kitaev2006}, intensive effort has been devoted
to find material realizations of such a state~\cite{Jackeli2009,WinterReview,Witczak-Krempa2014,Rau16}.
Promising candidates are the layered honeycomb iridates
{\nairo}~\cite{Choi2012} and
 $\alpha$-{\liiro}~\cite{Gretarsson2013,Freund2016}
 as well RuCl$_3$~\cite{Johnson2015,Banerjee2016, Banerjee2017,
Winter2017, Winter2018}. 
and the three-dimensional  lattices
$\beta$-{\liiro} and $\gamma$-{\liiro}~\cite{Takayama2015, BiffinBeta, BiffinGamma,Li2017, Modic2014}.
However, these materials order magnetically
 either in a zig-zag structure as in {\nairo} and RuCl$_3$ 
or in  incommensurate spiral structures as in the {\liiro} 
polymorphs~\cite{Williams2016}. 
 This magnetic long-range order has been 
attributed to the presence of further non-Kitaev interactions
~\cite{Katukuri2014,Rau2014,Winter2016,natalia2018,natalia2018b}
partly reminiscent of underlying delocalized 
quasimolecular orbitals~\cite{Mazin2012,Foyevtsova2013,Li2015}.
Attemps to modulate the magnetic interactions in terms of
pressure 
resulted in dimerized structures in the layered cases
with no sign
of spin liquid behavior~\cite{Hermann2018,Biesner2018,bastien2018}.

Recently, however,
a new member of this family was synthesized 
by substituting  in
$\alpha$-Li$_2$IrO$_3$ 
interlayer lithium ions  by protons ($^1$H$^+$)~\cite{o2012production,Bette2017,Kitagawa2018,Takayama2018}.
Measurements of magnetic susceptibility,
specific heat, and nuclear magnetic resonance (NMR) on the resulting {\hliiro} showed no sign
of magnetic order down to 0.05 K~\cite{Kitagawa2018}. Initial studies of the
deuterium variant suggest a similar response~\cite{Takayama2018}. However, the
low temperature heat capacity of {\hliiro} only accounts for a small $\sim$1 -
2\% of the total ln(2) spin entropy, suggesting that the majority of spin
excitations are gapped out~\cite{Kitagawa2018}. This can be contrasted with the
pure Kitaev model, for which thermally excited static fluxes produce a
pronounced low temperature peak accounting for 50\% of ln(2)
entropy~\cite{nasu2015thermal}. These deviations from the Kitaev model have
been interpreted in terms of a model with sizeable interlayer couplings
mediated by the H atoms~\cite{Kevin2017}. Alternatively, it has been suggested
that the low temperature scaling of the specific heat, susceptibility, and NMR
response may be explained by a small fraction of defect-induced local
moments~\cite{kimchi2018heat}. Attempts to estimate the magnetic interactions
via {\it ab initio} studies are currently lacking. 

In this work, we investigate the specific influence of hydrogen $^1$H$^+$ and deuterium (D$^+$ = $^2$H$^+$)
substitution on the
local magnetic interactions of (H/D)$_3$LiIr$_2$O$_6$ for both pristine and
structurally disordered samples. For that we perform extensive density functional theory simulations
combined with numerical model calculations.  For edge-sharing iridates, the magnetic interactions
are known to be highly sensitive to local structural details such as the
Ir-O-Ir bond angles~\cite{Jackeli2009, Chaloupka2010, Yamaji2014, Rau2014,
Katukuri2014, Winter2016}. Here, we show that the H-bond geometry similarly
strongly affects the magnetic Hamiltonian. This emphasizes the importance
of
determining the precise H/D positions, which are usually unavailable from
x-ray analysis. We therefore analyze
 the stability of the symmetrical O-H-O
bonds with respect to (i) zero-point fluctuations of the H positions,
(ii) deuteration and (iii) the presence of structural disorder. We
find that such bonds are likely to be unstable to both latter
perturbations (ii)-(iii), leading
to strongly modified interactions in deuterated or disordered samples. 

\begin{figure}[t]
  \includegraphics[width=\linewidth]{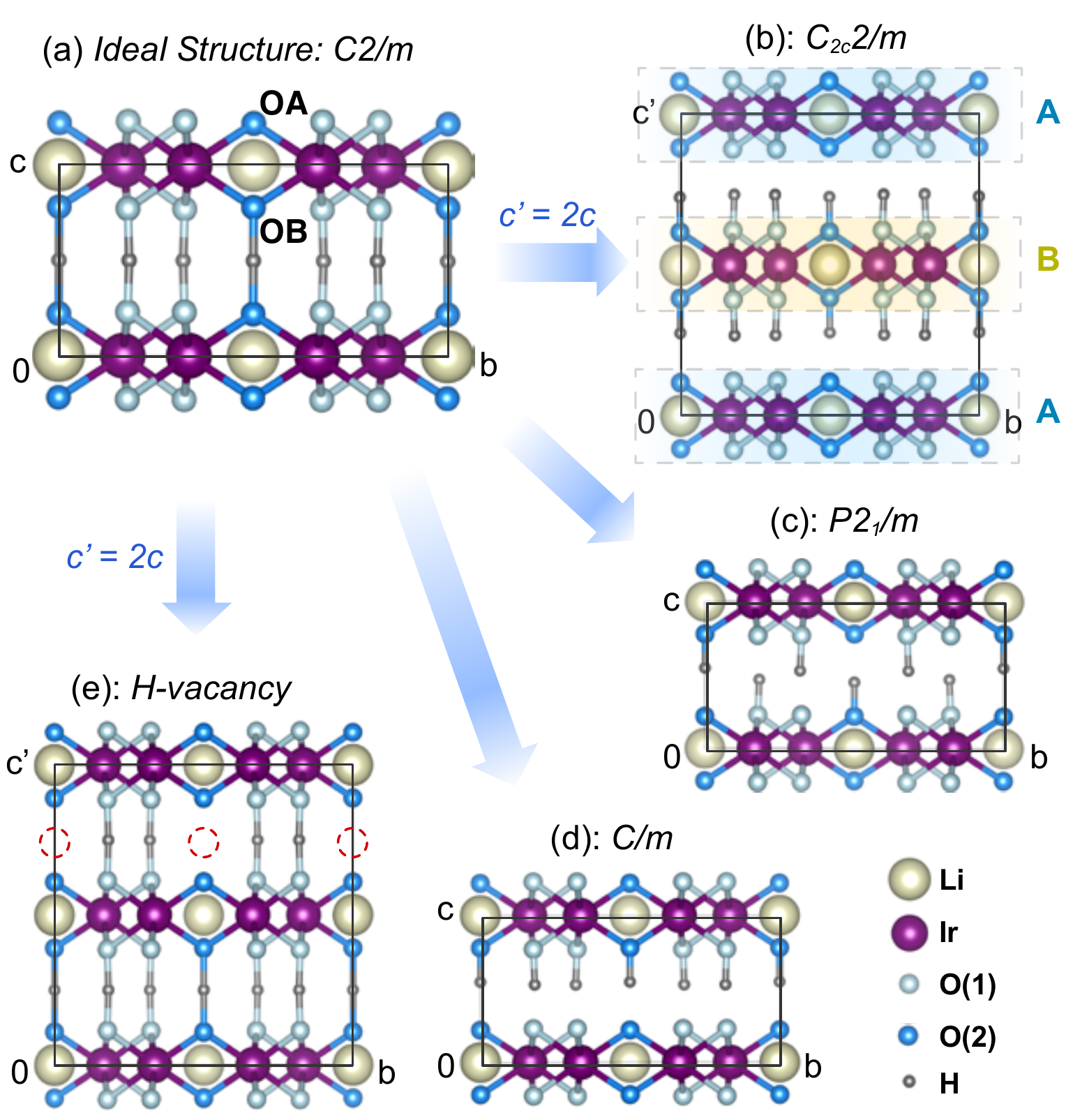}  
  \caption{Hypothetical structures considered to investigate the dependence of magnetic interactions on H-bond geometry. The two bridging ligand oxygen contributed to Ir-Ir hoppings along the Z-bonds are labeled as OA and OB. For structure (b) and (e), doubling of the $c$-axis leads to two inequivalent layers A and B.}
\label{fig:diffH}
\end{figure}

We first study the influence of the H-bond geometry on the electronic
structure and magnetic interactions. Starting from the ideal
structure~\cite{Bette2017}, the H coordinates were
relaxed with the Vienna {\it ab initio} simulation package (VASP)~\cite{Kresse1996,Hafner2008}
in the GGA approximation constraining the
symmetry to be: (a) $C2/m$, (b) $C_{2c}2/m$, (c) $P2_1/m$, and (d) $Cm$
(see Fig.~\ref{fig:diffH}).
For the ideal $C2/m$ structure (Fig.~\ref{fig:diffH} (a)), the symmetrical interlayer O-H-O bonds 
are maintained with minimal modification; the relaxed O(1)-H and
O(2)-H distances are $d_{O(1)H} = $ 1.23 {\AA} and $d_{O(2)H} = $ 1.27 {\AA},
respectively. In contrast, the lower symmetry structures (Fig.~\ref{fig:diffH} (b)-(d)) feature
various hypothetical patterns of asymmetric long/short O$\cdot\cdot\cdot$H-O
bonds. In each case, the short O-H distances are
in the range $d_{OH} =$ 1.05 - 1.10 {\AA}, while the longer O$\cdot\cdot\cdot$H
distances are $d_{OH}$ = 1.36 - 1.50 {\AA}. The hopping parameters of the
relaxed structures were obtained by full-potential linearized augmented
plane-wave (LAPW) calculations~\cite{Wien2k} within GGA.
The magnetic interactions were
estimated by 2-site exact diagonalization of the corresponding
tight-binding model with Hubbard, Hund's  and spin-orbit coupling 
interactions~\cite{Winter2016,Winter2017CT}, and are presented in Table~\ref{tab:mag}. Full
computational and structure details can be found in the Supplemental Material~\footnote{See Supplemental Material which contains structural parameters, density of states, and hopping parameters for various structures as well as Ref~\onlinecite{Wien2k, Bloechl1994, Perdew1996, Kresse1996, Hafner2008, Foyevtsova2013, Rau2014}}.

\begin{figure}[b]
  \includegraphics[width=\linewidth]{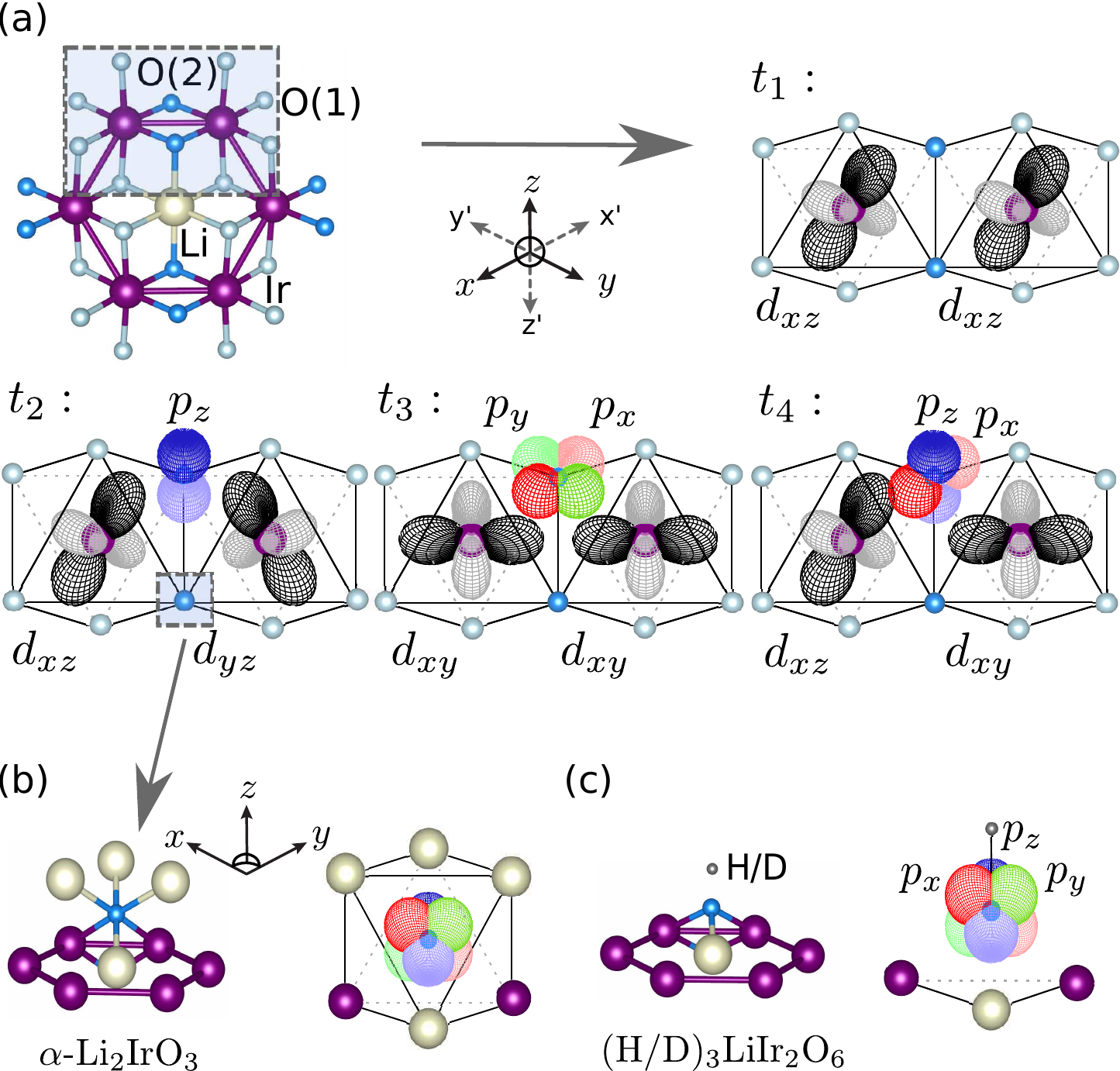}  
  \caption{(a) O $p$-orbital contributions to the effective Ir-Ir hopping integrals $t_1$-$t_4$ for the Z-bond. (b)-(c): Comparison of the oxygen coordination environments in $\alpha$-Li$_2$IrO$_3$ and (H/D)$_3$LiIr$_2$O$_6$, respectively.}
\label{fig:local}
\end{figure}

At the level of a single Ir, the $d$-orbital Wannier functions extend to the
neighboring O atoms. Modification of the oxygen bonding environment via
 H doping (delithiation in this case)  therefore affects both the effective Ir crystal field splitting,
and the $d$-$d$ hopping integrals. In the parent $\alpha$-Li$_2$IrO$_3$
material, the Li$^+$ and Ir$^{4+}$ ions form a pseudo-octahedral environment
around each O atom (Fig.~\ref{fig:local} (b)). In {\hliiro}, the O environment
becomes pseudo-tetrahedral, with the O-(H/D) bond vectors lying along the cubic
[111] direction (Fig.~\ref{fig:local}(c)).
Hybridization of the O $2p$ and H $1s$
orbitals produces an effective trigonal crystal field at the oxygen sites. The
resultant mixing of the O $p$-orbitals modifies the Ir $d$-orbital Wannier
functions in two key ways. First, the effective off-diagonal crystal field
terms are enhanced with increasing O-H hybridization. Second, oxygen mediated
$d$-$d$ hopping integrals involving single (multiple) O $p$-orbitals are
suppressed (enhanced). For example, within $C2/m$ symmetry, the $d$-$d$ hopping
integrals for the Z-bond can be written in terms of $t_1 =
t_{xz,xz} = t_{yz,yz}$, $t_2 = t_{xz,yz} = t_{yz,xz}$, $t_3 = t_{xy,xy}$, and
$t_4 = t_{xz,xy} = t_{yz,xy} = t_{xy,xz} = t_{xy,yz}$ 
(Fig.~\ref{fig:local} (a)).
Comparing the results
for the various structures, we find that $|t_2|$ is systematically reduced with
decreasing O-H distance, due to suppression of the hopping paths like
Ir($d_{xz}$)$ \to $O($p_z$)$ \to$Ir($ d_{yz}$). Similarly, $|t_3|$ and $|t_4|$
are enhanced through new hopping paths such as Ir($d_{xz}$)$ \to
$O($p_z$)$\to$H($s$)$\to$O($p_x$)$\to$Ir($ d_{xy}$). The hopping intergrals
for various structures are in the Supplemental Material \cite{Note1}.

The modification of the hopping integrals affects the magnetic couplings, which can be written
as $H_{\rm spin} = \sum_{\langle ij\rangle} \mathbf{S}_i \cdot \mathbf{J}_{ij} \cdot \mathbf{S}_j,$
where $\mathbf{J}_{ij}$ is a $3\times 3$ matrix
and $\langle ij\rangle$ denotes a sum over all nearest neighbor sites. Here $H_{\rm spin}$ is a spin model of interacting spin-orbit entangled pseudospins, rather than real spins. Within $C2/m$ symmetry, the interactions along the nearest-neighbor Z-bond (Fig.~1(a)) are described by a symmetric matrix:
\begin{align}
\mathbf{J}_{s} = \left(\begin{array}{ccc} J_1 & \Gamma_1 & \Gamma_1^{\prime} \\ \Gamma_1 & J_1 & \Gamma_1^{\prime} \\ \Gamma_1^{\prime} & \Gamma_1^{\prime} & J_1+K_1\end{array}\right),
\end{align}
For the ideal structure (Fig.~\ref{fig:diffH} (a)), we estimate 
  $J_1 = -1.3$, $K_1=-15.4$, $\Gamma_1 = +1.5$, and $\Gamma_1^\prime = -5.1$ meV for the Z-bond where positive (negative) signs correspond
to antiferromagnetic (ferromagnetic) interactions.
 For the lower symmetry X and Y bonds, additional constants are required to specify the interactions~\cite{Winter2016}. Averaging over these values yields $J_1 \approx -0.8$, $K_1\approx -18.9$, $\Gamma_1 \approx +0.3$, and $\Gamma_1^\prime \approx -5.4$ meV.  Compared to $\alpha$-Li$_2$IrO$_3$, the increased Ir-O-Ir bond angle of {\hliiro} suppresses $t_3$ and leads to the reduced $\Gamma_1$. The large $\Gamma_1^\prime$ values for the ideal structure of {\hliiro} are due to the enhanced crystal-field splitting.
Note that the computed interactions $(J,K,\Gamma,\Gamma^\prime) = (-1.3,-15.4,+1.5,-5.1)$ meV 
are given in the reference frame $xyz$ shown in Fig.~\ref{fig:local}. By  performing a 
$\pi$-rotation around the cubic (111) direction, as
outlined in Ref.~\onlinecite{chaloupka2015hidden}, the $xyz$ reference frame transforms to $x'y'z'$ (Fig.~\ref{fig:local})
and the above set of interactions
transform to $(J,K,\Gamma,\Gamma^\prime) = (-11.1, +13.9, -8.3,-0.2)$ meV. This model lies close to the classical tricritical
point~\cite{Rau2014} between ferromagnetic, zigzag, and incommensurate spiral
orders, which may strongly contribute to the observed lack of magnetic order.
 The interlayer interactions are anisotropic, but have
a magnitude $\lesssim 1.5$ meV, which is somewhat smaller than those considered
in the model of Ref.~\onlinecite{Kevin2017}. Inclusion of further neighbor
couplings (such as an antiferromagnetic third neighbor $J_3$) will tend to stabilize zigzag
correlations~\cite{Winter2016}, which is consistent with the differing $^1$H
and $^7$Li NMR Knight shift discussed in Ref.~\onlinecite{Kitagawa2018}. 

The specific effects of the O-H hybridization can
be clearly seen by comparing the above
computed interactions for the $C2/m$ structure (Fig.~\ref{fig:diffH} (a))
with symmetric O-H-O bonds to those for the various lower symmetry structures (Fig.~\ref{fig:diffH} (b)-(d)).
We will discuss the interactions in the $xyz$ reference frame of Fig.~\ref{fig:local}.
For structure Fig.~\ref{fig:diffH} (b), the doubled $c$-axis provides two distinct layers, with O-H
distances being equal within each layer. In the H-rich layer B, the reduced
$d_{OH} = 1.05$ {\AA} results in a significant suppression of $|t_2|$, which
reduces $K_1$ to $-7$ meV. In contrast, for the H-poor layer A, for which
$d_{OH} = 1.49$ \AA, we find an enhancement of $K_1$ to $-23.7$ meV.  For
structures Fig.~\ref{fig:diffH} (c)-(d), the symmetry of the two Ir-O-Ir hopping paths is
broken by the asymmetric hydrogen atom positions. This produces several
effects. First, the lower symmetry enhances $\Gamma_1^\prime$, and allows for a
large antisymmetric Dzyalloshinskii-Moriya interaction
$\mathbf{D}\cdot(\mathbf{S}_i\times \mathbf{S}_j)$ due to breaking of local
inversion symmetry. Second, the perfect balance of the hopping paths via the
two bridging oxygens is disrupted, leading to an enhancement of the Heisenberg
exchange $J_1$, and suppression of $K_1$. This trend is also followed in the
presence of the H-vacancies, which we have considered in terms of structure
Fig.~\ref{fig:diffH} (e), which is obtained from the ideal structure Fig.~\ref{fig:diffH} (a) by removing one of the H
atoms for the Z-bond. This further reduces the local symmetry, leading to large
$J_1$, $\Gamma_1^\prime$ and DM-interactions. Taken together, the large overall
variances in the computed interactions  demonstrate an extreme sensitivity to
the details of the H-bonds.
\begin{table}[t]
\caption {Nearest-neighbor Z-bond magnetic interactions in meV for {\hliiro} obtained by exact diagonalization on two-site cluster for different structures employing $U$ = 1.7 eV, $J_{\rm H}$ = 0.3 eV, $\lambda = 0.4$ eV and full crystal-field terms.}
\centering\def\arraystretch{1.1}
\label{tab:mag}
\begin{ruledtabular}
\begin{tabular}{lrrrrrr}
Structure &$J_1$ & $K_1$ &$\Gamma_1$ &$\Gamma_1^{\prime}$  & ($D_x$, $D_y$, $D_z$)\\
\hline
(a) Ideal ($C2/m$) & -1.3 & -15.4 & +1.5 & -5.1 &( -,-,- ) \\ 
\hline
(b) ($C_{2c}2/m$)\\
H-Poor (Layer A) & -0.3 & -23.7  &-1.5 & -3.3 &(-, -, -)\\ 
H-Rich (Layer B)& -2.7 &  -7.0 & +3.1 &-4.0 &(-, -, -)\\  
\hline 
(c) ($P2_1/m$) &-7.6 & -1.6 & +2.3 & -9.1 & (2.3, 2.3, 0.1)\\ 
\hline
(d) ($Cm$)& -12.0 & +6.8&+2.4 & -10.6 &(2.7, 2.7, 0.7)\\ 
\hline
(e) H-vacancy& -16.3 & +3.7 & +1.6 & -18.5 &$\pm$(6.1 6.1 -7.2)
\end{tabular}
\end{ruledtabular}
\end{table}

Given the sensitivity of the magnetic interactions, it is crucial to determine the precise positions of the H and D atoms.
Following Ref.~\onlinecite{mckenzie2014effect,Charles1997}, the interlayer O-H-O hydrogen bonds can be classified as strong, low-barrier, or weak, depending on the shape of the energy potential as a function of hydrogen position. If the distance between the two oxygens $d_{\rm OO}$ is short ($\lesssim 2.4$ \AA)~\cite{Schiott1998}, the bond is characterized either by a single-well potential or by a double-well potential with a barrier that is smaller than the vibrational zero-point energy $E_0$ (ZPE). In this case, the probability density $|\Psi_0^H(x)|^2$ for the H position in the vibrational ground state  is expected to display a single peak at the midpoint between the O atoms. This is considered a {\it strong} H-bond. 
For larger distance ($d_{\rm OO}$ $\sim 2.55 $ \AA), the barrier typically becomes of similar magnitude to the ZPE, leading to a double peak in $|\Psi_0^H(x)|^2$. The H rapidly tunnels between the two minima, with a characteristic frequency $\omega^*$ that can be estimated from the energy difference of the lowest two vibrational levels $\hbar\omega^* \sim (E_1 - E_0)$. Provided this tunnel splitting $\omega^*$ is large compared to the experimental time scales, the dynamical fluctuations of the H will be averaged out in measurements of the magnetic response. This is termed a {\it low-barrier} H-bond. Finally, for still larger distances ($d_{\rm OO}$ $\gtrsim 2.6$ \AA) the development of a large barrier suppresses tunnelling ($\omega^* \to 0$), leading to a {\it weak} H-bond. In this limit, the H-atoms will become increasingly localized into asymmetric O$\cdot \cdot \cdot$H-O bonds on experimental time-scales.

\begin{figure}
\includegraphics[width=\linewidth]{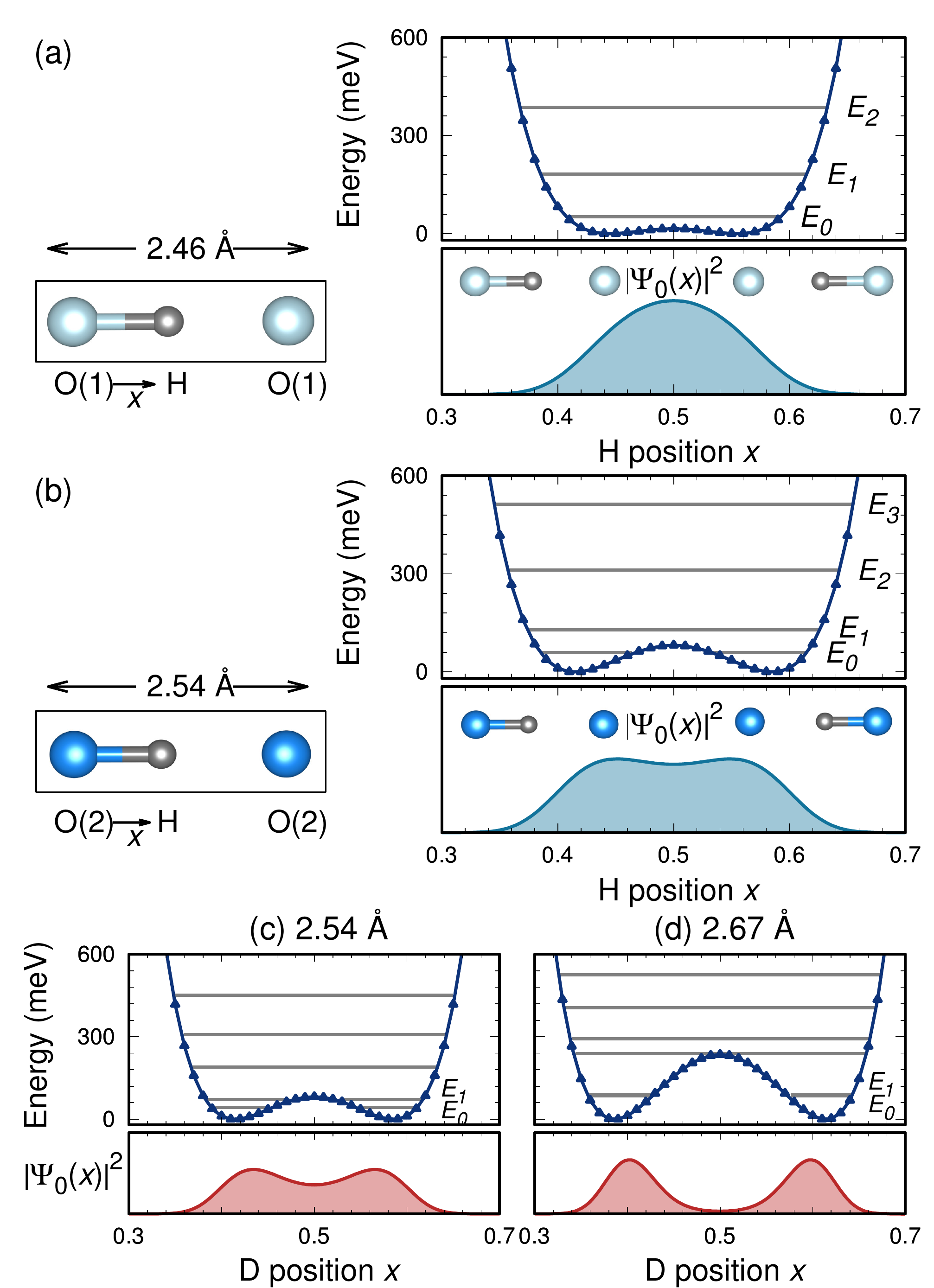}  
\caption{Energies of the crystal with O-H distance $d_{OH} = xd_{OO}$ for (a) O(1)-H-O(1) bond 1 and (b) O(2)-H-O(2) bond 2 when fixing another hydrogen in the middle of oxygen and oxygen. (c) Energies with replacing Hydrogen by Deuterium for bond 2, and (d) replacing Hydrogen by Deuterium for bond 2 with stretching of O-O distance by 5\%. $\{ E_n \}$ are the vibrational energies.}
\label{fig:ene}
\end{figure}

In order to investigate the H-bond potentials for {\hliiro}, we performed total
energy calculations as a function of hydrogen positions along the O-O vector
($d_{\rm OH} = xd_{\rm OO}$), starting from the the experimental $C2/m$
structure of Ref.~\onlinecite{Bette2017}. There are two kinds of O-H-O bonds:
bond 1 (O(1)-H-O(1)) with $d_{\rm OO}$ $\sim$ 2.46 {\AA}, and bond 2
(O(2)-H-O(2)) of $d_{\rm OO}$ $\sim$ 2.54 \AA.  As shown in Fig.~\ref{fig:ene}
(a)-(b), the potential curves for both bond types have two local minima. To
estimate the ZPE, we fit the obtained potential energy curves with the form
$V(x) = -Ax^2 + Bx^4$ and computed the vibrational eigenstates for the
corresponding Hamiltonian $H = -\frac{\hbar^2}{2m}\nabla^2 + V (x)$. The
vibrational energies $\{ E_n \}$ are indicated in Fig.~\ref{fig:ene}. For bond
1, the ZPE ($E_0$) exceeds the barrier energy, suggesting classification as a
strong H-bond. For bond 2, the ZPE lies below the barrier, leading to a double
peak in $|\Psi_0^H(x)|^2$. However, the tunnel splitting remains large
$\hbar\omega^* = E_1 - E_0 \sim 70$ meV compared to the magnetic interactions,
suggesting classification as a low-barrier H-bond. 
 On this basis, in pristine samples of
{\hliiro}, the symmetry of the O-H-O bonds should be maintained on time scales
relevant to the magnetic response. The effective magnetic interactions should
reflect those computed above 
for the ideal $C2/m$ structure Fig.~\ref{fig:diffH} (a).
 
This conclusion, however, does not hold for D$_3$LiIr$_2$O$_6$. In similar
layered structures, D-substitution tends to increase the interlayer O-O
distance, e.g. on the order of 5\% for
(H/D)CrO$_2$.~\cite{christensen1977isotope} At the same time, the larger mass
of D significantly reduces the ZPE (see Fig.~\ref{fig:ene} (d)). As a result,
the propensity for deuterated systems to form asymmetric O$\cdot\cdot\cdot$D-O
bonds is strongly enhanced~\cite{matsuo2000proton}. For example, while the
O-H-O bonds in HCrO$_2$ remain symmetric, DCrO$_2$ undergoes a structural
transition at $T_c \sim 320$ K, occasioned by the formation of asymmetric bonds
in the bulk~\cite{matsuo2006isotope,dolin2007study}. Similar
effects~\cite{ueda2014hydrogen,yamamoto2016theoretical} are also observed in
the deuterated organic spin-liquid candidate $\kappa$-D$_3$-(Cat-EDT-TTF)$_2$
($T_c \sim 185$ K). We therefore computed the deuterium vibrational energies
for D$_3$LiIr$_2$O$_6$ including a O-O distance of 2.54 {\AA} (see
Fig.~\ref{fig:ene} (c)), and a $\sim$5\% stretch of O-O distances to 2.67 {\AA}
(see Fig.~\ref{fig:ene} (d)) for bond 2. The tunnel splittings of $\hbar
\omega^* \sim$ 30 meV in Fig.~\ref{fig:ene} (c), and $\sim$ 3 meV in
Fig.~\ref{fig:ene} (d) suggest the formation of asymmetric
{O-D$\cdot\cdot\cdot$D} bonds in D$_3$LiIr$_2$O$_6$ samples on magnetically
relevant time scales. Given the similarity of the magnetic and
structural energy scales, non-trivial effects of their coupling
may also appear. However, the bulk magnetic interactions should differ
strongly from pristine H$_3$LiIr$_2$O$_6$, reflecting those computed for e.g.
structures Fig.~\ref{fig:diffH} (b)-(d).

The question therefore remains whether the observed suppression of magnetic order in (H/D)$_{3}$LiIr$_{2}$O$_6$ results primarily from frustration (i.e.~quantum fluctuations), or from disorder-induced inhomogeneity~\cite{kimchi2018heat,Kimchi2018Yb,Willans2010}. Experimentally, if both H and D systems display similar low temperature response despite differing degrees of H/D localization, it could be taken as evidence that both are dominated by disorder. Since H/D atoms near grain boundaries and crystalline defects will likely form conventional short O-H bonds, disorder effects on the interactions may be enhanced.

 It is therefore relevant to consider possible sources of structural disorder.
One source could be related to the H substitution through 
the  delithiation method Li$_2$MO$_3$ $\to$ H$_x$Li$_{2-2y-x}$MO$_{3-y}$
where the resulting compounds are generally thought to be rather disordered with variable stoichiometry ($x,y$)~\cite{paik2002lithium}.
Inhomogeneity may also arise in stoichiometric samples related to the microstructure and distribution of H
atoms. 
Due to the differing coordination environment around the O atoms, introduction of H requires the shifting of the hexagonal layers with respect
to each
other. The resultant shear stresses have been observed to cleave the crystals along the $ab$-plane, producing thin platelets with
numerous stacking faults~\cite{tang2000preparation,weber2017trivalent}, increasing substantially the sample surface area.
For isolated thin platelets, the formation of short O-H bonds at the surface would lead to strongly different interactions from the bulk.
Furthermore, complete coverage of {\it both} the top and bottom surfaces of a
platelet with H$^+$ would require a surplus of protons. A finite concentration
of hydrogen vacancies in the sample is therefore required for charge neutrality, modifying the local interactions (i.e. structure Fig.~\ref{fig:diffH} (e)).
In view of the above considerations, the magnetic
response is likely be sensitive to both the stoichiometry and microstructure of
the samples, through the strong influence of the H-bonds on the local
interactions.

In summary, we have investigated how the introduction of hydrogen affects the
structure and magnetic properties of (H/D)$_3$LiIr$_2$O$_6$. The H-bond
geometry strongly affects the local magnetic interactions due to hybridization
with the bridging oxygen ligands. For the H-system, we conclude that the
interlayer O-H-O bonds are likely to remain symmetrical in bulk pristine
samples due to large zero-point fluctuations of the H positions. In this case,
the estimated magnetic model lies close to the classical tricritical point
between ferromagnetic, zigzag and incommensurate spiral orders. This strongly hints
to a lack of magnetic order in the quantum case. Upon deuteration
or in the presence of structural defects and grain boundaries, the symmetrical
O-(H/D)-O bonds are destabilized, strongly modifying the local magnetic
couplings. Considering previous studies, such defects may naturally arise
through the method employed in the preparation of
(H/D)$_3$LiIr$_2$O$_6$. Given the strong sensitivity of the magnetic
Hamiltonian to the H-bond geometry, further studies of the specific
microstructure and composition of the samples will provide significant insight
into their magnetic response. 
\begin{acknowledgments}
We acknowledge useful discussions with Sananda Biswas, Vladislav Borisov, George Jackeli,
Tomo Takayama  and Hidenori Takagi. We acknowledge support by the Deutsche Forschungsgemeinschaft through grant SFB/TR 49 and the computer time was allotted by
the Centre for Scientific Computing (CSC) in Frankfurt.
\end{acknowledgments}  
\bibliography{ref}

\begin{thebibliography}{61}%
\makeatletter
\providecommand \@ifxundefined [1]{%
 \@ifx{#1\undefined}
}%
\providecommand \@ifnum [1]{%
 \ifnum #1\expandafter \@firstoftwo
 \else \expandafter \@secondoftwo
 \fi
}%
\providecommand \@ifx [1]{%
 \ifx #1\expandafter \@firstoftwo
 \else \expandafter \@secondoftwo
 \fi
}%
\providecommand \natexlab [1]{#1}%
\providecommand \enquote  [1]{``#1''}%
\providecommand \bibnamefont  [1]{#1}%
\providecommand \bibfnamefont [1]{#1}%
\providecommand \citenamefont [1]{#1}%
\providecommand \href@noop [0]{\@secondoftwo}%
\providecommand \href [0]{\begingroup \@sanitize@url \@href}%
\providecommand \@href[1]{\@@startlink{#1}\@@href}%
\providecommand \@@href[1]{\endgroup#1\@@endlink}%
\providecommand \@sanitize@url [0]{\catcode `\\12\catcode `\$12\catcode
  `\&12\catcode `\#12\catcode `\^12\catcode `\_12\catcode `\%12\relax}%
\providecommand \@@startlink[1]{}%
\providecommand \@@endlink[0]{}%
\providecommand \url  [0]{\begingroup\@sanitize@url \@url }%
\providecommand \@url [1]{\endgroup\@href {#1}{\urlprefix }}%
\providecommand \urlprefix  [0]{URL }%
\providecommand \Eprint [0]{\href }%
\providecommand \doibase [0]{http://dx.doi.org/}%
\providecommand \selectlanguage [0]{\@gobble}%
\providecommand \bibinfo  [0]{\@secondoftwo}%
\providecommand \bibfield  [0]{\@secondoftwo}%
\providecommand \translation [1]{[#1]}%
\providecommand \BibitemOpen [0]{}%
\providecommand \bibitemStop [0]{}%
\providecommand \bibitemNoStop [0]{.\EOS\space}%
\providecommand \EOS [0]{\spacefactor3000\relax}%
\providecommand \BibitemShut  [1]{\csname bibitem#1\endcsname}%
\let\auto@bib@innerbib\@empty
\bibitem [{\citenamefont {Kitaev}(2006)}]{Kitaev2006}%
  \BibitemOpen
  \bibfield  {author} {\bibinfo {author} {\bibfnamefont {A.}~\bibnamefont
  {Kitaev}},\ }\href@noop {} {\bibfield  {journal} {\bibinfo  {journal} {Annals
  of Physics}\ }\textbf {\bibinfo {volume} {321}} (\bibinfo {year}
  {2006})}\BibitemShut {NoStop}%
\bibitem [{\citenamefont {Jackeli}\ and\ \citenamefont
  {Khaliullin}(2009)}]{Jackeli2009}%
  \BibitemOpen
  \bibfield  {author} {\bibinfo {author} {\bibfnamefont {G.}~\bibnamefont
  {Jackeli}}\ and\ \bibinfo {author} {\bibfnamefont {G.}~\bibnamefont
  {Khaliullin}},\ }\href@noop {} {\bibfield  {journal} {\bibinfo  {journal}
  {Phys. Rev. Lett.}\ }\textbf {\bibinfo {volume} {102}},\ \bibinfo {pages}
  {017205} (\bibinfo {year} {2009})}\BibitemShut {NoStop}%
\bibitem [{\citenamefont {Winter}\ \emph
  {et~al.}(2017{\natexlab{a}})\citenamefont {Winter}, \citenamefont {Tsirlin},
  \citenamefont {Daghofer}, \citenamefont {van~den Brink}, \citenamefont
  {Singh}, \citenamefont {Gegenwart},\ and\ \citenamefont
  {Valent\'{\i}}}]{WinterReview}%
  \BibitemOpen
  \bibfield  {author} {\bibinfo {author} {\bibfnamefont {S.~M.}\ \bibnamefont
  {Winter}}, \bibinfo {author} {\bibfnamefont {A.~A.}\ \bibnamefont {Tsirlin}},
  \bibinfo {author} {\bibfnamefont {M.}~\bibnamefont {Daghofer}}, \bibinfo
  {author} {\bibfnamefont {J.}~\bibnamefont {van~den Brink}}, \bibinfo {author}
  {\bibfnamefont {Y.}~\bibnamefont {Singh}}, \bibinfo {author} {\bibfnamefont
  {P.}~\bibnamefont {Gegenwart}}, \ and\ \bibinfo {author} {\bibfnamefont
  {R.}~\bibnamefont {Valent\'{\i}}},\ }\href@noop {} {\bibfield  {journal}
  {\bibinfo  {journal} {Journal of Physics: Condensed Matter}\ }\textbf
  {\bibinfo {volume} {29}},\ \bibinfo {pages} {493002} (\bibinfo {year}
  {2017}{\natexlab{a}})}\BibitemShut {NoStop}%
\bibitem [{\citenamefont {Witczak-Krempa}\ \emph {et~al.}(2014)\citenamefont
  {Witczak-Krempa}, \citenamefont {Chen}, \citenamefont {Kim},\ and\
  \citenamefont {Balents}}]{Witczak-Krempa2014}%
  \BibitemOpen
  \bibfield  {author} {\bibinfo {author} {\bibfnamefont {W.}~\bibnamefont
  {Witczak-Krempa}}, \bibinfo {author} {\bibfnamefont {G.}~\bibnamefont
  {Chen}}, \bibinfo {author} {\bibfnamefont {Y.~B.}\ \bibnamefont {Kim}}, \
  and\ \bibinfo {author} {\bibfnamefont {L.}~\bibnamefont {Balents}},\
  }\href@noop {} {\bibfield  {journal} {\bibinfo  {journal} {Annual Review of
  Condensed Matter Physics}\ }\textbf {\bibinfo {volume} {5}},\ \bibinfo
  {pages} {57} (\bibinfo {year} {2014})}\BibitemShut {NoStop}%
\bibitem [{\citenamefont {Rau}\ \emph {et~al.}(2016)\citenamefont {Rau},
  \citenamefont {Lee},\ and\ \citenamefont {Kee}}]{Rau16}%
  \BibitemOpen
  \bibfield  {author} {\bibinfo {author} {\bibfnamefont {J.~G.}\ \bibnamefont
  {Rau}}, \bibinfo {author} {\bibfnamefont {E.~K.-H.}\ \bibnamefont {Lee}}, \
  and\ \bibinfo {author} {\bibfnamefont {H.-Y.}\ \bibnamefont {Kee}},\
  }\href@noop {} {\bibfield  {journal} {\bibinfo  {journal} {Annual Review of
  Condensed Matter Physics}\ }\textbf {\bibinfo {volume} {7}},\ \bibinfo
  {pages} {195} (\bibinfo {year} {2016})}\BibitemShut {NoStop}%
\bibitem [{\citenamefont {Choi}\ \emph {et~al.}(2012)\citenamefont {Choi},
  \citenamefont {Coldea}, \citenamefont {Kolmogorov}, \citenamefont
  {Lancaster}, \citenamefont {Mazin}, \citenamefont {Blundell}, \citenamefont
  {Radaelli}, \citenamefont {Singh}, \citenamefont {Gegenwart}, \citenamefont
  {Choi}, \citenamefont {Cheong}, \citenamefont {Baker}, \citenamefont
  {Stock},\ and\ \citenamefont {Taylor}}]{Choi2012}%
  \BibitemOpen
  \bibfield  {author} {\bibinfo {author} {\bibfnamefont {S.~K.}\ \bibnamefont
  {Choi}}, \bibinfo {author} {\bibfnamefont {R.}~\bibnamefont {Coldea}},
  \bibinfo {author} {\bibfnamefont {A.~N.}\ \bibnamefont {Kolmogorov}},
  \bibinfo {author} {\bibfnamefont {T.}~\bibnamefont {Lancaster}}, \bibinfo
  {author} {\bibfnamefont {I.~I.}\ \bibnamefont {Mazin}}, \bibinfo {author}
  {\bibfnamefont {S.~J.}\ \bibnamefont {Blundell}}, \bibinfo {author}
  {\bibfnamefont {P.~G.}\ \bibnamefont {Radaelli}}, \bibinfo {author}
  {\bibfnamefont {Y.}~\bibnamefont {Singh}}, \bibinfo {author} {\bibfnamefont
  {P.}~\bibnamefont {Gegenwart}}, \bibinfo {author} {\bibfnamefont {K.~R.}\
  \bibnamefont {Choi}}, \bibinfo {author} {\bibfnamefont {S.-W.}\ \bibnamefont
  {Cheong}}, \bibinfo {author} {\bibfnamefont {P.~J.}\ \bibnamefont {Baker}},
  \bibinfo {author} {\bibfnamefont {C.}~\bibnamefont {Stock}}, \ and\ \bibinfo
  {author} {\bibfnamefont {J.}~\bibnamefont {Taylor}},\ }\href@noop {}
  {\bibfield  {journal} {\bibinfo  {journal} {Phys. Rev. Lett.}\ }\textbf
  {\bibinfo {volume} {108}},\ \bibinfo {pages} {127204} (\bibinfo {year}
  {2012})}\BibitemShut {NoStop}%
\bibitem [{\citenamefont {Gretarsson}\ \emph {et~al.}(2013)\citenamefont
  {Gretarsson}, \citenamefont {Clancy}, \citenamefont {Liu}, \citenamefont
  {Hill}, \citenamefont {Bozin}, \citenamefont {Singh}, \citenamefont {Manni},
  \citenamefont {Gegenwart}, \citenamefont {Kim}, \citenamefont {Said},
  \citenamefont {Casa}, \citenamefont {Gog}, \citenamefont {Upton},
  \citenamefont {Kim}, \citenamefont {Yu}, \citenamefont {Katukuri},
  \citenamefont {Hozoi}, \citenamefont {van~den Brink},\ and\ \citenamefont
  {Kim}}]{Gretarsson2013}%
  \BibitemOpen
  \bibfield  {author} {\bibinfo {author} {\bibfnamefont {H.}~\bibnamefont
  {Gretarsson}}, \bibinfo {author} {\bibfnamefont {J.~P.}\ \bibnamefont
  {Clancy}}, \bibinfo {author} {\bibfnamefont {X.}~\bibnamefont {Liu}},
  \bibinfo {author} {\bibfnamefont {J.~P.}\ \bibnamefont {Hill}}, \bibinfo
  {author} {\bibfnamefont {E.}~\bibnamefont {Bozin}}, \bibinfo {author}
  {\bibfnamefont {Y.}~\bibnamefont {Singh}}, \bibinfo {author} {\bibfnamefont
  {S.}~\bibnamefont {Manni}}, \bibinfo {author} {\bibfnamefont
  {P.}~\bibnamefont {Gegenwart}}, \bibinfo {author} {\bibfnamefont
  {J.}~\bibnamefont {Kim}}, \bibinfo {author} {\bibfnamefont {A.~H.}\
  \bibnamefont {Said}}, \bibinfo {author} {\bibfnamefont {D.}~\bibnamefont
  {Casa}}, \bibinfo {author} {\bibfnamefont {T.}~\bibnamefont {Gog}}, \bibinfo
  {author} {\bibfnamefont {M.~H.}\ \bibnamefont {Upton}}, \bibinfo {author}
  {\bibfnamefont {H.-S.}\ \bibnamefont {Kim}}, \bibinfo {author} {\bibfnamefont
  {J.}~\bibnamefont {Yu}}, \bibinfo {author} {\bibfnamefont {V.~M.}\
  \bibnamefont {Katukuri}}, \bibinfo {author} {\bibfnamefont {L.}~\bibnamefont
  {Hozoi}}, \bibinfo {author} {\bibfnamefont {J.}~\bibnamefont {van~den
  Brink}}, \ and\ \bibinfo {author} {\bibfnamefont {Y.-J.}\ \bibnamefont
  {Kim}},\ }\href@noop {} {\bibfield  {journal} {\bibinfo  {journal} {Phys.
  Rev. Lett.}\ }\textbf {\bibinfo {volume} {110}},\ \bibinfo {pages} {076402}
  (\bibinfo {year} {2013})}\BibitemShut {NoStop}%
\bibitem [{\citenamefont {Freund}\ \emph {et~al.}(2016)\citenamefont {Freund},
  \citenamefont {Williams}, \citenamefont {Johnson}, \citenamefont {Coldea},
  \citenamefont {Gegenwart},\ and\ \citenamefont {Jesche}}]{Freund2016}%
  \BibitemOpen
  \bibfield  {author} {\bibinfo {author} {\bibfnamefont {F.}~\bibnamefont
  {Freund}}, \bibinfo {author} {\bibfnamefont {S.~C.}\ \bibnamefont
  {Williams}}, \bibinfo {author} {\bibfnamefont {R.~D.}\ \bibnamefont
  {Johnson}}, \bibinfo {author} {\bibfnamefont {R.}~\bibnamefont {Coldea}},
  \bibinfo {author} {\bibfnamefont {P.}~\bibnamefont {Gegenwart}}, \ and\
  \bibinfo {author} {\bibfnamefont {A.}~\bibnamefont {Jesche}},\ }\href@noop {}
  {\bibfield  {journal} {\bibinfo  {journal} {Scientific Reports}\ }\textbf
  {\bibinfo {volume} {6}},\ \bibinfo {pages} {35362} (\bibinfo {year}
  {2016})}\BibitemShut {NoStop}%
\bibitem [{\citenamefont {Johnson}\ \emph {et~al.}(2015)\citenamefont
  {Johnson}, \citenamefont {Williams}, \citenamefont {Haghighirad},
  \citenamefont {Singleton}, \citenamefont {Zapf}, \citenamefont {Manuel},
  \citenamefont {Mazin}, \citenamefont {Li}, \citenamefont {Jeschke},
  \citenamefont {Valent\'{\i}},\ and\ \citenamefont {Coldea}}]{Johnson2015}%
  \BibitemOpen
  \bibfield  {author} {\bibinfo {author} {\bibfnamefont {R.~D.}\ \bibnamefont
  {Johnson}}, \bibinfo {author} {\bibfnamefont {S.~C.}\ \bibnamefont
  {Williams}}, \bibinfo {author} {\bibfnamefont {A.~A.}\ \bibnamefont
  {Haghighirad}}, \bibinfo {author} {\bibfnamefont {J.}~\bibnamefont
  {Singleton}}, \bibinfo {author} {\bibfnamefont {V.}~\bibnamefont {Zapf}},
  \bibinfo {author} {\bibfnamefont {P.}~\bibnamefont {Manuel}}, \bibinfo
  {author} {\bibfnamefont {I.~I.}\ \bibnamefont {Mazin}}, \bibinfo {author}
  {\bibfnamefont {Y.}~\bibnamefont {Li}}, \bibinfo {author} {\bibfnamefont
  {H.~O.}\ \bibnamefont {Jeschke}}, \bibinfo {author} {\bibfnamefont
  {R.}~\bibnamefont {Valent\'{\i}}}, \ and\ \bibinfo {author} {\bibfnamefont
  {R.}~\bibnamefont {Coldea}},\ }\href@noop {} {\bibfield  {journal} {\bibinfo
  {journal} {Phys. Rev. B}\ }\textbf {\bibinfo {volume} {92}},\ \bibinfo
  {pages} {235119} (\bibinfo {year} {2015})}\BibitemShut {NoStop}%
\bibitem [{\citenamefont {Banerjee}\ \emph {et~al.}(2016)\citenamefont
  {Banerjee}, \citenamefont {Bridges}, \citenamefont {Yan}, \citenamefont
  {Aczel}, \citenamefont {Li}, \citenamefont {Stone}, \citenamefont {Granroth},
  \citenamefont {Lumsden}, \citenamefont {Yiu}, \citenamefont {Knolle},
  \citenamefont {Bhattacharjee}, \citenamefont {Kovrizhin}, \citenamefont
  {Moessner}, \citenamefont {Tennant}, \citenamefont {Mandrus},\ and\
  \citenamefont {Nagler}}]{Banerjee2016}%
  \BibitemOpen
  \bibfield  {author} {\bibinfo {author} {\bibfnamefont {A.}~\bibnamefont
  {Banerjee}}, \bibinfo {author} {\bibfnamefont {C.~A.}\ \bibnamefont
  {Bridges}}, \bibinfo {author} {\bibfnamefont {J.-Q.}\ \bibnamefont {Yan}},
  \bibinfo {author} {\bibfnamefont {A.~A.}\ \bibnamefont {Aczel}}, \bibinfo
  {author} {\bibfnamefont {L.}~\bibnamefont {Li}}, \bibinfo {author}
  {\bibfnamefont {M.~B.}\ \bibnamefont {Stone}}, \bibinfo {author}
  {\bibfnamefont {G.~E.}\ \bibnamefont {Granroth}}, \bibinfo {author}
  {\bibfnamefont {M.~D.}\ \bibnamefont {Lumsden}}, \bibinfo {author}
  {\bibfnamefont {Y.}~\bibnamefont {Yiu}}, \bibinfo {author} {\bibfnamefont
  {J.}~\bibnamefont {Knolle}}, \bibinfo {author} {\bibfnamefont
  {S.}~\bibnamefont {Bhattacharjee}}, \bibinfo {author} {\bibfnamefont {D.~L.}\
  \bibnamefont {Kovrizhin}}, \bibinfo {author} {\bibfnamefont {R.}~\bibnamefont
  {Moessner}}, \bibinfo {author} {\bibfnamefont {D.~A.}\ \bibnamefont
  {Tennant}}, \bibinfo {author} {\bibfnamefont {D.~G.}\ \bibnamefont
  {Mandrus}}, \ and\ \bibinfo {author} {\bibfnamefont {S.~E.}\ \bibnamefont
  {Nagler}},\ }\href@noop {} {\bibfield  {journal} {\bibinfo  {journal} {Nature
  Materials}\ }\textbf {\bibinfo {volume} {15}},\ \bibinfo {pages} {733}
  (\bibinfo {year} {2016})}\BibitemShut {NoStop}%
\bibitem [{\citenamefont {Banerjee}\ \emph {et~al.}(2017)\citenamefont
  {Banerjee}, \citenamefont {Yan}, \citenamefont {Knolle}, \citenamefont
  {Bridges}, \citenamefont {Stone}, \citenamefont {Lumsden}, \citenamefont
  {Mandrus}, \citenamefont {Tennant}, \citenamefont {Moessner},\ and\
  \citenamefont {Nagler}}]{Banerjee2017}%
  \BibitemOpen
  \bibfield  {author} {\bibinfo {author} {\bibfnamefont {A.}~\bibnamefont
  {Banerjee}}, \bibinfo {author} {\bibfnamefont {J.}~\bibnamefont {Yan}},
  \bibinfo {author} {\bibfnamefont {J.}~\bibnamefont {Knolle}}, \bibinfo
  {author} {\bibfnamefont {C.~A.}\ \bibnamefont {Bridges}}, \bibinfo {author}
  {\bibfnamefont {M.~B.}\ \bibnamefont {Stone}}, \bibinfo {author}
  {\bibfnamefont {M.~D.}\ \bibnamefont {Lumsden}}, \bibinfo {author}
  {\bibfnamefont {D.~G.}\ \bibnamefont {Mandrus}}, \bibinfo {author}
  {\bibfnamefont {D.~A.}\ \bibnamefont {Tennant}}, \bibinfo {author}
  {\bibfnamefont {R.}~\bibnamefont {Moessner}}, \ and\ \bibinfo {author}
  {\bibfnamefont {S.~E.}\ \bibnamefont {Nagler}},\ }\href@noop {} {\bibfield
  {journal} {\bibinfo  {journal} {Science}\ }\textbf {\bibinfo {volume}
  {356}},\ \bibinfo {pages} {1055} (\bibinfo {year} {2017})}\BibitemShut
  {NoStop}%
\bibitem [{\citenamefont {Winter}\ \emph
  {et~al.}(2017{\natexlab{b}})\citenamefont {Winter}, \citenamefont {Riedl},
  \citenamefont {Maksimov}, \citenamefont {Chernyshev}, \citenamefont
  {Honecker},\ and\ \citenamefont {Valent\'{\i}}}]{Winter2017}%
  \BibitemOpen
  \bibfield  {author} {\bibinfo {author} {\bibfnamefont {S.~M.}\ \bibnamefont
  {Winter}}, \bibinfo {author} {\bibfnamefont {K.}~\bibnamefont {Riedl}},
  \bibinfo {author} {\bibfnamefont {P.~A.}\ \bibnamefont {Maksimov}}, \bibinfo
  {author} {\bibfnamefont {A.~L.}\ \bibnamefont {Chernyshev}}, \bibinfo
  {author} {\bibfnamefont {A.}~\bibnamefont {Honecker}}, \ and\ \bibinfo
  {author} {\bibfnamefont {R.}~\bibnamefont {Valent\'{\i}}},\ }\href@noop {}
  {\bibfield  {journal} {\bibinfo  {journal} {Nature Communications}\ }\textbf
  {\bibinfo {volume} {8}},\ \bibinfo {pages} {1152} (\bibinfo {year}
  {2017}{\natexlab{b}})}\BibitemShut {NoStop}%
\bibitem [{\citenamefont {Winter}\ \emph {et~al.}(2018)\citenamefont {Winter},
  \citenamefont {Riedl}, \citenamefont {Kaib}, \citenamefont {Coldea},\ and\
  \citenamefont {Valent\'{\i}}}]{Winter2018}%
  \BibitemOpen
  \bibfield  {author} {\bibinfo {author} {\bibfnamefont {S.~M.}\ \bibnamefont
  {Winter}}, \bibinfo {author} {\bibfnamefont {K.}~\bibnamefont {Riedl}},
  \bibinfo {author} {\bibfnamefont {D.}~\bibnamefont {Kaib}}, \bibinfo {author}
  {\bibfnamefont {R.}~\bibnamefont {Coldea}}, \ and\ \bibinfo {author}
  {\bibfnamefont {R.}~\bibnamefont {Valent\'{\i}}},\ }\href@noop {} {\bibfield
  {journal} {\bibinfo  {journal} {Phys. Rev. Lett.}\ }\textbf {\bibinfo
  {volume} {120}},\ \bibinfo {pages} {077203} (\bibinfo {year}
  {2018})}\BibitemShut {NoStop}%
\bibitem [{\citenamefont {Takayama}\ \emph {et~al.}(2015)\citenamefont
  {Takayama}, \citenamefont {Kato}, \citenamefont {Dinnebier}, \citenamefont
  {Nuss}, \citenamefont {Kono}, \citenamefont {Veiga}, \citenamefont {Fabbris},
  \citenamefont {Haskel},\ and\ \citenamefont {Takagi}}]{Takayama2015}%
  \BibitemOpen
  \bibfield  {author} {\bibinfo {author} {\bibfnamefont {T.}~\bibnamefont
  {Takayama}}, \bibinfo {author} {\bibfnamefont {A.}~\bibnamefont {Kato}},
  \bibinfo {author} {\bibfnamefont {R.}~\bibnamefont {Dinnebier}}, \bibinfo
  {author} {\bibfnamefont {J.}~\bibnamefont {Nuss}}, \bibinfo {author}
  {\bibfnamefont {H.}~\bibnamefont {Kono}}, \bibinfo {author} {\bibfnamefont
  {L.~S.~I.}\ \bibnamefont {Veiga}}, \bibinfo {author} {\bibfnamefont
  {G.}~\bibnamefont {Fabbris}}, \bibinfo {author} {\bibfnamefont
  {D.}~\bibnamefont {Haskel}}, \ and\ \bibinfo {author} {\bibfnamefont
  {H.}~\bibnamefont {Takagi}},\ }\href@noop {} {\bibfield  {journal} {\bibinfo
  {journal} {Phys. Rev. Lett.}\ }\textbf {\bibinfo {volume} {114}},\ \bibinfo
  {pages} {077202} (\bibinfo {year} {2015})}\BibitemShut {NoStop}%
\bibitem [{\citenamefont {Biffin}\ \emph
  {et~al.}(2014{\natexlab{a}})\citenamefont {Biffin}, \citenamefont {Johnson},
  \citenamefont {Choi}, \citenamefont {Freund}, \citenamefont {Manni},
  \citenamefont {Bombardi}, \citenamefont {Manuel}, \citenamefont {Gegenwart},\
  and\ \citenamefont {Coldea}}]{BiffinBeta}%
  \BibitemOpen
  \bibfield  {author} {\bibinfo {author} {\bibfnamefont {A.}~\bibnamefont
  {Biffin}}, \bibinfo {author} {\bibfnamefont {R.~D.}\ \bibnamefont {Johnson}},
  \bibinfo {author} {\bibfnamefont {S.}~\bibnamefont {Choi}}, \bibinfo {author}
  {\bibfnamefont {F.}~\bibnamefont {Freund}}, \bibinfo {author} {\bibfnamefont
  {S.}~\bibnamefont {Manni}}, \bibinfo {author} {\bibfnamefont
  {A.}~\bibnamefont {Bombardi}}, \bibinfo {author} {\bibfnamefont
  {P.}~\bibnamefont {Manuel}}, \bibinfo {author} {\bibfnamefont
  {P.}~\bibnamefont {Gegenwart}}, \ and\ \bibinfo {author} {\bibfnamefont
  {R.}~\bibnamefont {Coldea}},\ }\href@noop {} {\bibfield  {journal} {\bibinfo
  {journal} {Phys. Rev. B}\ }\textbf {\bibinfo {volume} {90}},\ \bibinfo
  {pages} {205116} (\bibinfo {year} {2014}{\natexlab{a}})}\BibitemShut
  {NoStop}%
\bibitem [{\citenamefont {Biffin}\ \emph
  {et~al.}(2014{\natexlab{b}})\citenamefont {Biffin}, \citenamefont {Johnson},
  \citenamefont {Kimchi}, \citenamefont {Morris}, \citenamefont {Bombardi},
  \citenamefont {Analytis}, \citenamefont {Vishwanath},\ and\ \citenamefont
  {Coldea}}]{BiffinGamma}%
  \BibitemOpen
  \bibfield  {author} {\bibinfo {author} {\bibfnamefont {A.}~\bibnamefont
  {Biffin}}, \bibinfo {author} {\bibfnamefont {R.~D.}\ \bibnamefont {Johnson}},
  \bibinfo {author} {\bibfnamefont {I.}~\bibnamefont {Kimchi}}, \bibinfo
  {author} {\bibfnamefont {R.}~\bibnamefont {Morris}}, \bibinfo {author}
  {\bibfnamefont {A.}~\bibnamefont {Bombardi}}, \bibinfo {author}
  {\bibfnamefont {J.~G.}\ \bibnamefont {Analytis}}, \bibinfo {author}
  {\bibfnamefont {A.}~\bibnamefont {Vishwanath}}, \ and\ \bibinfo {author}
  {\bibfnamefont {R.}~\bibnamefont {Coldea}},\ }\href@noop {} {\bibfield
  {journal} {\bibinfo  {journal} {Phys. Rev. Lett.}\ }\textbf {\bibinfo
  {volume} {113}},\ \bibinfo {pages} {197201} (\bibinfo {year}
  {2014}{\natexlab{b}})}\BibitemShut {NoStop}%
\bibitem [{\citenamefont {Li}\ \emph {et~al.}(2017)\citenamefont {Li},
  \citenamefont {Winter}, \citenamefont {Jeschke},\ and\ \citenamefont
  {Valent\'{\i}}}]{Li2017}%
  \BibitemOpen
  \bibfield  {author} {\bibinfo {author} {\bibfnamefont {Y.}~\bibnamefont
  {Li}}, \bibinfo {author} {\bibfnamefont {S.~M.}\ \bibnamefont {Winter}},
  \bibinfo {author} {\bibfnamefont {H.~O.}\ \bibnamefont {Jeschke}}, \ and\
  \bibinfo {author} {\bibfnamefont {R.}~\bibnamefont {Valent\'{\i}}},\
  }\href@noop {} {\bibfield  {journal} {\bibinfo  {journal} {Phys. Rev. B}\
  }\textbf {\bibinfo {volume} {95}},\ \bibinfo {pages} {045129} (\bibinfo
  {year} {2017})}\BibitemShut {NoStop}%
\bibitem [{\citenamefont {Modic}\ \emph {et~al.}(2014)\citenamefont {Modic},
  \citenamefont {Smidt}, \citenamefont {Kimchi}, \citenamefont {Breznay},
  \citenamefont {Biffin}, \citenamefont {Choi}, \citenamefont {Johnson},
  \citenamefont {Coldea}, \citenamefont {Watkins-Curry},\ and\ \citenamefont
  {McCandless}}]{Modic2014}%
  \BibitemOpen
  \bibfield  {author} {\bibinfo {author} {\bibfnamefont {K.~A.}\ \bibnamefont
  {Modic}}, \bibinfo {author} {\bibfnamefont {T.~E.}\ \bibnamefont {Smidt}},
  \bibinfo {author} {\bibfnamefont {I.}~\bibnamefont {Kimchi}}, \bibinfo
  {author} {\bibfnamefont {N.~P.}\ \bibnamefont {Breznay}}, \bibinfo {author}
  {\bibfnamefont {A.}~\bibnamefont {Biffin}}, \bibinfo {author} {\bibfnamefont
  {S.}~\bibnamefont {Choi}}, \bibinfo {author} {\bibfnamefont {R.~D.}\
  \bibnamefont {Johnson}}, \bibinfo {author} {\bibfnamefont {R.}~\bibnamefont
  {Coldea}}, \bibinfo {author} {\bibfnamefont {P.}~\bibnamefont
  {Watkins-Curry}}, \ and\ \bibinfo {author} {\bibfnamefont {G.~T.}\
  \bibnamefont {McCandless}},\ }\href@noop {} {\bibfield  {journal} {\bibinfo
  {journal} {Nat. Commun.}\ }\textbf {\bibinfo {volume} {5}},\ \bibinfo {pages}
  {4203} (\bibinfo {year} {2014})}\BibitemShut {NoStop}%
\bibitem [{\citenamefont {Williams}\ \emph {et~al.}(2016)\citenamefont
  {Williams}, \citenamefont {Johnson}, \citenamefont {Freund}, \citenamefont
  {Choi}, \citenamefont {Jesche}, \citenamefont {Kimchi}, \citenamefont
  {Manni}, \citenamefont {Bombardi}, \citenamefont {Manuel}, \citenamefont
  {Gegenwart},\ and\ \citenamefont {Coldea}}]{Williams2016}%
  \BibitemOpen
  \bibfield  {author} {\bibinfo {author} {\bibfnamefont {S.~C.}\ \bibnamefont
  {Williams}}, \bibinfo {author} {\bibfnamefont {R.~D.}\ \bibnamefont
  {Johnson}}, \bibinfo {author} {\bibfnamefont {F.}~\bibnamefont {Freund}},
  \bibinfo {author} {\bibfnamefont {S.}~\bibnamefont {Choi}}, \bibinfo {author}
  {\bibfnamefont {A.}~\bibnamefont {Jesche}}, \bibinfo {author} {\bibfnamefont
  {I.}~\bibnamefont {Kimchi}}, \bibinfo {author} {\bibfnamefont
  {S.}~\bibnamefont {Manni}}, \bibinfo {author} {\bibfnamefont
  {A.}~\bibnamefont {Bombardi}}, \bibinfo {author} {\bibfnamefont
  {P.}~\bibnamefont {Manuel}}, \bibinfo {author} {\bibfnamefont
  {P.}~\bibnamefont {Gegenwart}}, \ and\ \bibinfo {author} {\bibfnamefont
  {R.}~\bibnamefont {Coldea}},\ }\href@noop {} {\bibfield  {journal} {\bibinfo
  {journal} {Phys. Rev. B}\ }\textbf {\bibinfo {volume} {93}},\ \bibinfo
  {pages} {195158} (\bibinfo {year} {2016})}\BibitemShut {NoStop}%
\bibitem [{\citenamefont {Katukuri}\ \emph {et~al.}(2014)\citenamefont
  {Katukuri}, \citenamefont {Nishimoto}, \citenamefont {Yushankhai},
  \citenamefont {Stoyanova}, \citenamefont {Kandpal}, \citenamefont {Choi},
  \citenamefont {Coldea}, \citenamefont {Rousochatzakis}, \citenamefont
  {Hozoi},\ and\ \citenamefont {van~den Brink}}]{Katukuri2014}%
  \BibitemOpen
  \bibfield  {author} {\bibinfo {author} {\bibfnamefont {V.~M.}\ \bibnamefont
  {Katukuri}}, \bibinfo {author} {\bibfnamefont {S.}~\bibnamefont {Nishimoto}},
  \bibinfo {author} {\bibfnamefont {V.}~\bibnamefont {Yushankhai}}, \bibinfo
  {author} {\bibfnamefont {A.}~\bibnamefont {Stoyanova}}, \bibinfo {author}
  {\bibfnamefont {H.}~\bibnamefont {Kandpal}}, \bibinfo {author} {\bibfnamefont
  {S.}~\bibnamefont {Choi}}, \bibinfo {author} {\bibfnamefont {R.}~\bibnamefont
  {Coldea}}, \bibinfo {author} {\bibfnamefont {I.}~\bibnamefont
  {Rousochatzakis}}, \bibinfo {author} {\bibfnamefont {L.}~\bibnamefont
  {Hozoi}}, \ and\ \bibinfo {author} {\bibfnamefont {J.}~\bibnamefont {van~den
  Brink}},\ }\href@noop {} {\bibfield  {journal} {\bibinfo  {journal} {New
  Journal of Physics}\ }\textbf {\bibinfo {volume} {16}},\ \bibinfo {pages}
  {013056} (\bibinfo {year} {2014})}\BibitemShut {NoStop}%
\bibitem [{\citenamefont {Rau}\ \emph {et~al.}(2014)\citenamefont {Rau},
  \citenamefont {Lee},\ and\ \citenamefont {Kee}}]{Rau2014}%
  \BibitemOpen
  \bibfield  {author} {\bibinfo {author} {\bibfnamefont {J.~G.}\ \bibnamefont
  {Rau}}, \bibinfo {author} {\bibfnamefont {E.~K.-H.}\ \bibnamefont {Lee}}, \
  and\ \bibinfo {author} {\bibfnamefont {H.-Y.}\ \bibnamefont {Kee}},\
  }\href@noop {} {\bibfield  {journal} {\bibinfo  {journal} {Phys. Rev. Lett.}\
  }\textbf {\bibinfo {volume} {112}},\ \bibinfo {pages} {077204} (\bibinfo
  {year} {2014})}\BibitemShut {NoStop}%
\bibitem [{\citenamefont {Winter}\ \emph {et~al.}(2016)\citenamefont {Winter},
  \citenamefont {Li}, \citenamefont {Jeschke},\ and\ \citenamefont
  {Valent\'{\i}}}]{Winter2016}%
  \BibitemOpen
  \bibfield  {author} {\bibinfo {author} {\bibfnamefont {S.~M.}\ \bibnamefont
  {Winter}}, \bibinfo {author} {\bibfnamefont {Y.}~\bibnamefont {Li}}, \bibinfo
  {author} {\bibfnamefont {H.~O.}\ \bibnamefont {Jeschke}}, \ and\ \bibinfo
  {author} {\bibfnamefont {R.}~\bibnamefont {Valent\'{\i}}},\ }\href@noop {}
  {\bibfield  {journal} {\bibinfo  {journal} {Phys. Rev. B}\ }\textbf {\bibinfo
  {volume} {93}},\ \bibinfo {pages} {214431} (\bibinfo {year}
  {2016})}\BibitemShut {NoStop}%
\bibitem [{\citenamefont {Ducatman}\ \emph {et~al.}(2018)\citenamefont
  {Ducatman}, \citenamefont {Rousochatzakis},\ and\ \citenamefont
  {Perkins}}]{natalia2018}%
  \BibitemOpen
  \bibfield  {author} {\bibinfo {author} {\bibfnamefont {S.}~\bibnamefont
  {Ducatman}}, \bibinfo {author} {\bibfnamefont {I.}~\bibnamefont
  {Rousochatzakis}}, \ and\ \bibinfo {author} {\bibfnamefont {N.~B.}\
  \bibnamefont {Perkins}},\ }\href@noop {} {\bibfield  {journal} {\bibinfo
  {journal} {Physical Review B}\ }\textbf {\bibinfo {volume} {97}},\ \bibinfo
  {pages} {125125} (\bibinfo {year} {2018})}\BibitemShut {NoStop}%
\bibitem [{\citenamefont {Rousochatzakis}\ and\ \citenamefont
  {Perkins}(2018)}]{natalia2018b}%
  \BibitemOpen
  \bibfield  {author} {\bibinfo {author} {\bibfnamefont {I.}~\bibnamefont
  {Rousochatzakis}}\ and\ \bibinfo {author} {\bibfnamefont {N.~B.}\
  \bibnamefont {Perkins}},\ }\href@noop {} {\bibfield  {journal} {\bibinfo
  {journal} {Physical Review B}\ }\textbf {\bibinfo {volume} {97}},\ \bibinfo
  {pages} {174423} (\bibinfo {year} {2018})}\BibitemShut {NoStop}%
\bibitem [{\citenamefont {Mazin}\ \emph {et~al.}(2012)\citenamefont {Mazin},
  \citenamefont {Jeschke}, \citenamefont {Foyevtsova}, \citenamefont
  {Valent\'{\i}},\ and\ \citenamefont {Khomskii}}]{Mazin2012}%
  \BibitemOpen
  \bibfield  {author} {\bibinfo {author} {\bibfnamefont {I.~I.}\ \bibnamefont
  {Mazin}}, \bibinfo {author} {\bibfnamefont {H.~O.}\ \bibnamefont {Jeschke}},
  \bibinfo {author} {\bibfnamefont {K.}~\bibnamefont {Foyevtsova}}, \bibinfo
  {author} {\bibfnamefont {R.}~\bibnamefont {Valent\'{\i}}}, \ and\ \bibinfo
  {author} {\bibfnamefont {D.~I.}\ \bibnamefont {Khomskii}},\ }\href@noop {}
  {\bibfield  {journal} {\bibinfo  {journal} {Phys. Rev. Lett.}\ }\textbf
  {\bibinfo {volume} {109}},\ \bibinfo {pages} {197201} (\bibinfo {year}
  {2012})}\BibitemShut {NoStop}%
\bibitem [{\citenamefont {Foyevtsova}\ \emph {et~al.}(2013)\citenamefont
  {Foyevtsova}, \citenamefont {Jeschke}, \citenamefont {Mazin}, \citenamefont
  {Khomskii},\ and\ \citenamefont {Valent\'{\i}}}]{Foyevtsova2013}%
  \BibitemOpen
  \bibfield  {author} {\bibinfo {author} {\bibfnamefont {K.}~\bibnamefont
  {Foyevtsova}}, \bibinfo {author} {\bibfnamefont {H.~O.}\ \bibnamefont
  {Jeschke}}, \bibinfo {author} {\bibfnamefont {I.~I.}\ \bibnamefont {Mazin}},
  \bibinfo {author} {\bibfnamefont {D.~I.}\ \bibnamefont {Khomskii}}, \ and\
  \bibinfo {author} {\bibfnamefont {R.}~\bibnamefont {Valent\'{\i}}},\
  }\href@noop {} {\bibfield  {journal} {\bibinfo  {journal} {Phys. Rev. B}\
  }\textbf {\bibinfo {volume} {88}},\ \bibinfo {pages} {035107} (\bibinfo
  {year} {2013})}\BibitemShut {NoStop}%
\bibitem [{\citenamefont {Li}\ \emph {et~al.}(2015)\citenamefont {Li},
  \citenamefont {Foyevtsova}, \citenamefont {Jeschke},\ and\ \citenamefont
  {Valent\'{\i}}}]{Li2015}%
  \BibitemOpen
  \bibfield  {author} {\bibinfo {author} {\bibfnamefont {Y.}~\bibnamefont
  {Li}}, \bibinfo {author} {\bibfnamefont {K.}~\bibnamefont {Foyevtsova}},
  \bibinfo {author} {\bibfnamefont {H.~O.}\ \bibnamefont {Jeschke}}, \ and\
  \bibinfo {author} {\bibfnamefont {R.}~\bibnamefont {Valent\'{\i}}},\
  }\href@noop {} {\bibfield  {journal} {\bibinfo  {journal} {Phys. Rev. B}\
  }\textbf {\bibinfo {volume} {91}},\ \bibinfo {pages} {161101} (\bibinfo
  {year} {2015})}\BibitemShut {NoStop}%
\bibitem [{\citenamefont {Hermann}\ \emph {et~al.}(2018)\citenamefont
  {Hermann}, \citenamefont {Altmeyer}, \citenamefont {Ebad-Allah},
  \citenamefont {Freund}, \citenamefont {Jesche}, \citenamefont {Tsirlin},
  \citenamefont {Hanfland}, \citenamefont {Gegenwart}, \citenamefont {Mazin},
  \citenamefont {Khomskii}, \citenamefont {Valent\'{\i}},\ and\ \citenamefont
  {Kuntscher}}]{Hermann2018}%
  \BibitemOpen
  \bibfield  {author} {\bibinfo {author} {\bibfnamefont {V.}~\bibnamefont
  {Hermann}}, \bibinfo {author} {\bibfnamefont {M.}~\bibnamefont {Altmeyer}},
  \bibinfo {author} {\bibfnamefont {J.}~\bibnamefont {Ebad-Allah}}, \bibinfo
  {author} {\bibfnamefont {F.}~\bibnamefont {Freund}}, \bibinfo {author}
  {\bibfnamefont {A.}~\bibnamefont {Jesche}}, \bibinfo {author} {\bibfnamefont
  {A.~A.}\ \bibnamefont {Tsirlin}}, \bibinfo {author} {\bibfnamefont
  {M.}~\bibnamefont {Hanfland}}, \bibinfo {author} {\bibfnamefont
  {P.}~\bibnamefont {Gegenwart}}, \bibinfo {author} {\bibfnamefont {I.~I.}\
  \bibnamefont {Mazin}}, \bibinfo {author} {\bibfnamefont {D.~I.}\ \bibnamefont
  {Khomskii}}, \bibinfo {author} {\bibfnamefont {R.}~\bibnamefont
  {Valent\'{\i}}}, \ and\ \bibinfo {author} {\bibfnamefont {C.~A.}\
  \bibnamefont {Kuntscher}},\ }\href@noop {} {\bibfield  {journal} {\bibinfo
  {journal} {Phys. Rev. B}\ }\textbf {\bibinfo {volume} {97}},\ \bibinfo
  {pages} {020104} (\bibinfo {year} {2018})}\BibitemShut {NoStop}%
\bibitem [{\citenamefont {Biesner}\ \emph {et~al.}(2018)\citenamefont
  {Biesner}, \citenamefont {Biswas}, \citenamefont {Li}, \citenamefont {Saito},
  \citenamefont {Pustogow}, \citenamefont {Altmeyer}, \citenamefont {Wolter},
  \citenamefont {B\"uchner}, \citenamefont {Roslova}, \citenamefont {Doert},
  \citenamefont {Winter}, \citenamefont {Valent\'{\i}},\ and\ \citenamefont
  {Dressel}}]{Biesner2018}%
  \BibitemOpen
  \bibfield  {author} {\bibinfo {author} {\bibfnamefont {T.}~\bibnamefont
  {Biesner}}, \bibinfo {author} {\bibfnamefont {S.}~\bibnamefont {Biswas}},
  \bibinfo {author} {\bibfnamefont {W.}~\bibnamefont {Li}}, \bibinfo {author}
  {\bibfnamefont {Y.}~\bibnamefont {Saito}}, \bibinfo {author} {\bibfnamefont
  {A.}~\bibnamefont {Pustogow}}, \bibinfo {author} {\bibfnamefont
  {M.}~\bibnamefont {Altmeyer}}, \bibinfo {author} {\bibfnamefont {A.~U.~B.}\
  \bibnamefont {Wolter}}, \bibinfo {author} {\bibfnamefont {B.}~\bibnamefont
  {B\"uchner}}, \bibinfo {author} {\bibfnamefont {M.}~\bibnamefont {Roslova}},
  \bibinfo {author} {\bibfnamefont {T.}~\bibnamefont {Doert}}, \bibinfo
  {author} {\bibfnamefont {S.~M.}\ \bibnamefont {Winter}}, \bibinfo {author}
  {\bibfnamefont {R.}~\bibnamefont {Valent\'{\i}}}, \ and\ \bibinfo {author}
  {\bibfnamefont {M.}~\bibnamefont {Dressel}},\ }\href@noop {} {\bibfield
  {journal} {\bibinfo  {journal} {Phys. Rev. B}\ }\textbf {\bibinfo {volume}
  {97}},\ \bibinfo {pages} {220401} (\bibinfo {year} {2018})}\BibitemShut
  {NoStop}%
\bibitem [{\citenamefont {Bastien}\ \emph {et~al.}(2018)\citenamefont
  {Bastien}, \citenamefont {Garbarino}, \citenamefont {Yadav}, \citenamefont
  {Martinez-Casado}, \citenamefont {Rodr{\'\i}guez}, \citenamefont {Stahl},
  \citenamefont {Kusch}, \citenamefont {Limandri}, \citenamefont {Ray},
  \citenamefont {Lampen-Kelley} \emph {et~al.}}]{bastien2018}%
  \BibitemOpen
  \bibfield  {author} {\bibinfo {author} {\bibfnamefont {G.}~\bibnamefont
  {Bastien}}, \bibinfo {author} {\bibfnamefont {G.}~\bibnamefont {Garbarino}},
  \bibinfo {author} {\bibfnamefont {R.}~\bibnamefont {Yadav}}, \bibinfo
  {author} {\bibfnamefont {F.}~\bibnamefont {Martinez-Casado}}, \bibinfo
  {author} {\bibfnamefont {R.~B.}\ \bibnamefont {Rodr{\'\i}guez}}, \bibinfo
  {author} {\bibfnamefont {Q.}~\bibnamefont {Stahl}}, \bibinfo {author}
  {\bibfnamefont {M.}~\bibnamefont {Kusch}}, \bibinfo {author} {\bibfnamefont
  {S.}~\bibnamefont {Limandri}}, \bibinfo {author} {\bibfnamefont
  {R.}~\bibnamefont {Ray}}, \bibinfo {author} {\bibfnamefont {P.}~\bibnamefont
  {Lampen-Kelley}},  \emph {et~al.},\ }\href@noop {} {\bibfield  {journal}
  {\bibinfo  {journal} {Physical Review B}\ }\textbf {\bibinfo {volume} {97}},\
  \bibinfo {pages} {241108} (\bibinfo {year} {2018})}\BibitemShut {NoStop}%
\bibitem [{\citenamefont {O'Malley}\ \emph {et~al.}(2012)\citenamefont
  {O'Malley}, \citenamefont {Woodward},\ and\ \citenamefont
  {Verweij}}]{o2012production}%
  \BibitemOpen
  \bibfield  {author} {\bibinfo {author} {\bibfnamefont {M.~J.}\ \bibnamefont
  {O'Malley}}, \bibinfo {author} {\bibfnamefont {P.~M.}\ \bibnamefont
  {Woodward}}, \ and\ \bibinfo {author} {\bibfnamefont {H.}~\bibnamefont
  {Verweij}},\ }\href@noop {} {\bibfield  {journal} {\bibinfo  {journal}
  {Journal of Materials Chemistry}\ }\textbf {\bibinfo {volume} {22}},\
  \bibinfo {pages} {7782} (\bibinfo {year} {2012})}\BibitemShut {NoStop}%
\bibitem [{\citenamefont {Bette}\ \emph {et~al.}(2017)\citenamefont {Bette},
  \citenamefont {Takayama}, \citenamefont {Kitagawa}, \citenamefont {Takano},
  \citenamefont {Takagi},\ and\ \citenamefont {Dinnebier}}]{Bette2017}%
  \BibitemOpen
  \bibfield  {author} {\bibinfo {author} {\bibfnamefont {S.}~\bibnamefont
  {Bette}}, \bibinfo {author} {\bibfnamefont {T.}~\bibnamefont {Takayama}},
  \bibinfo {author} {\bibfnamefont {K.}~\bibnamefont {Kitagawa}}, \bibinfo
  {author} {\bibfnamefont {R.}~\bibnamefont {Takano}}, \bibinfo {author}
  {\bibfnamefont {H.}~\bibnamefont {Takagi}}, \ and\ \bibinfo {author}
  {\bibfnamefont {R.~E.}\ \bibnamefont {Dinnebier}},\ }\href@noop {} {\bibfield
   {journal} {\bibinfo  {journal} {Dalton Trans.}\ }\textbf {\bibinfo {volume}
  {46}},\ \bibinfo {pages} {15216} (\bibinfo {year} {2017})}\BibitemShut
  {NoStop}%
\bibitem [{\citenamefont {Kitagawa}\ \emph {et~al.}(2018)\citenamefont
  {Kitagawa}, \citenamefont {Takayama}, \citenamefont {Matsumoto},
  \citenamefont {Kato}, \citenamefont {Takano}, \citenamefont {Kishimoto},
  \citenamefont {Bette}, \citenamefont {Dinnebier}, \citenamefont {Jackeli},\
  and\ \citenamefont {Takagi}}]{Kitagawa2018}%
  \BibitemOpen
  \bibfield  {author} {\bibinfo {author} {\bibfnamefont {K.}~\bibnamefont
  {Kitagawa}}, \bibinfo {author} {\bibfnamefont {T.}~\bibnamefont {Takayama}},
  \bibinfo {author} {\bibfnamefont {Y.}~\bibnamefont {Matsumoto}}, \bibinfo
  {author} {\bibfnamefont {A.}~\bibnamefont {Kato}}, \bibinfo {author}
  {\bibfnamefont {R.}~\bibnamefont {Takano}}, \bibinfo {author} {\bibfnamefont
  {Y.}~\bibnamefont {Kishimoto}}, \bibinfo {author} {\bibfnamefont
  {S.}~\bibnamefont {Bette}}, \bibinfo {author} {\bibfnamefont
  {R.}~\bibnamefont {Dinnebier}}, \bibinfo {author} {\bibfnamefont
  {G.}~\bibnamefont {Jackeli}}, \ and\ \bibinfo {author} {\bibfnamefont
  {H.}~\bibnamefont {Takagi}},\ }\href@noop {} {\bibfield  {journal} {\bibinfo
  {journal} {Nature}\ }\textbf {\bibinfo {volume} {554}},\ \bibinfo {pages}
  {341} (\bibinfo {year} {2018})}\BibitemShut {NoStop}%
\bibitem [{\citenamefont {Takayama}()}]{Takayama2018}%
  \BibitemOpen
  \bibfield  {author} {\bibinfo {author} {\bibfnamefont {T.}~\bibnamefont
  {Takayama}},\ }\href@noop {} {\bibinfo  {journal} {APS March Meeting 2018}\
  }\BibitemShut {NoStop}%
\bibitem [{\citenamefont {Nasu}\ \emph {et~al.}(2015)\citenamefont {Nasu},
  \citenamefont {Udagawa},\ and\ \citenamefont {Motome}}]{nasu2015thermal}%
  \BibitemOpen
\bibfield  {journal} {  }\bibfield  {author} {\bibinfo {author} {\bibfnamefont
  {J.}~\bibnamefont {Nasu}}, \bibinfo {author} {\bibfnamefont {M.}~\bibnamefont
  {Udagawa}}, \ and\ \bibinfo {author} {\bibfnamefont {Y.}~\bibnamefont
  {Motome}},\ }\href@noop {} {\bibfield  {journal} {\bibinfo  {journal}
  {Physical Review B}\ }\textbf {\bibinfo {volume} {92}},\ \bibinfo {pages}
  {115122} (\bibinfo {year} {2015})}\BibitemShut {NoStop}%
\bibitem [{\citenamefont {Slagle}\ \emph {et~al.}(2018)\citenamefont {Slagle},
  \citenamefont {Choi}, \citenamefont {Chern},\ and\ \citenamefont
  {Kim}}]{Kevin2017}%
  \BibitemOpen
  \bibfield  {author} {\bibinfo {author} {\bibfnamefont {K.}~\bibnamefont
  {Slagle}}, \bibinfo {author} {\bibfnamefont {W.}~\bibnamefont {Choi}},
  \bibinfo {author} {\bibfnamefont {L.~E.}\ \bibnamefont {Chern}}, \ and\
  \bibinfo {author} {\bibfnamefont {Y.~B.}\ \bibnamefont {Kim}},\ }\href@noop
  {} {\bibfield  {journal} {\bibinfo  {journal} {Phys. Rev. B}\ }\textbf
  {\bibinfo {volume} {97}},\ \bibinfo {pages} {115159} (\bibinfo {year}
  {2018})}\BibitemShut {NoStop}%
\bibitem [{\citenamefont {Kimchi}\ \emph
  {et~al.}(2018{\natexlab{a}})\citenamefont {Kimchi}, \citenamefont
  {Sheckelton}, \citenamefont {McQueen},\ and\ \citenamefont
  {Lee}}]{kimchi2018heat}%
  \BibitemOpen
  \bibfield  {author} {\bibinfo {author} {\bibfnamefont {I.}~\bibnamefont
  {Kimchi}}, \bibinfo {author} {\bibfnamefont {J.~P.}\ \bibnamefont
  {Sheckelton}}, \bibinfo {author} {\bibfnamefont {T.~M.}\ \bibnamefont
  {McQueen}}, \ and\ \bibinfo {author} {\bibfnamefont {P.~A.}\ \bibnamefont
  {Lee}},\ }\href@noop {} {\bibfield  {journal} {\bibinfo  {journal}
  {arXiv:1803.00013}\ } (\bibinfo {year} {2018}{\natexlab{a}})}\BibitemShut
  {NoStop}%
\bibitem [{\citenamefont {Chaloupka}\ \emph {et~al.}(2010)\citenamefont
  {Chaloupka}, \citenamefont {Jackeli},\ and\ \citenamefont
  {Khaliullin}}]{Chaloupka2010}%
  \BibitemOpen
  \bibfield  {author} {\bibinfo {author} {\bibfnamefont {J.}~\bibnamefont
  {Chaloupka}}, \bibinfo {author} {\bibfnamefont {G.}~\bibnamefont {Jackeli}},
  \ and\ \bibinfo {author} {\bibfnamefont {G.}~\bibnamefont {Khaliullin}},\
  }\href@noop {} {\bibfield  {journal} {\bibinfo  {journal} {Phys. Rev. Lett.}\
  }\textbf {\bibinfo {volume} {105}},\ \bibinfo {pages} {027204} (\bibinfo
  {year} {2010})}\BibitemShut {NoStop}%
\bibitem [{\citenamefont {Yamaji}\ \emph {et~al.}(2014)\citenamefont {Yamaji},
  \citenamefont {Nomura}, \citenamefont {Kurita}, \citenamefont {Arita},\ and\
  \citenamefont {Imada}}]{Yamaji2014}%
  \BibitemOpen
  \bibfield  {author} {\bibinfo {author} {\bibfnamefont {Y.}~\bibnamefont
  {Yamaji}}, \bibinfo {author} {\bibfnamefont {Y.}~\bibnamefont {Nomura}},
  \bibinfo {author} {\bibfnamefont {M.}~\bibnamefont {Kurita}}, \bibinfo
  {author} {\bibfnamefont {R.}~\bibnamefont {Arita}}, \ and\ \bibinfo {author}
  {\bibfnamefont {M.}~\bibnamefont {Imada}},\ }\href@noop {} {\bibfield
  {journal} {\bibinfo  {journal} {Phys. Rev. Lett.}\ }\textbf {\bibinfo
  {volume} {113}},\ \bibinfo {pages} {107201} (\bibinfo {year}
  {2014})}\BibitemShut {NoStop}%
\bibitem [{\citenamefont {Kresse}\ and\ \citenamefont
  {Furthm\"{u}ller}(1996)}]{Kresse1996}%
  \BibitemOpen
  \bibfield  {author} {\bibinfo {author} {\bibfnamefont {G.}~\bibnamefont
  {Kresse}}\ and\ \bibinfo {author} {\bibfnamefont {J.}~\bibnamefont
  {Furthm\"{u}ller}},\ }\href@noop {} {\bibfield  {journal} {\bibinfo
  {journal} {Comput. Mater. Sci.}\ }\textbf {\bibinfo {volume} {6}},\ \bibinfo
  {pages} {15} (\bibinfo {year} {1996})}\BibitemShut {NoStop}%
\bibitem [{\citenamefont {Hafner}(2008)}]{Hafner2008}%
  \BibitemOpen
  \bibfield  {author} {\bibinfo {author} {\bibfnamefont {J.}~\bibnamefont
  {Hafner}},\ }\href@noop {} {\bibfield  {journal} {\bibinfo  {journal} {J.
  Comput. Chem.}\ }\textbf {\bibinfo {volume} {29}},\ \bibinfo {pages} {2044}
  (\bibinfo {year} {2008})}\BibitemShut {NoStop}%
\bibitem [{\citenamefont {Blaha}\ \emph {et~al.}(2001)\citenamefont {Blaha},
  \citenamefont {Schwarz}, \citenamefont {Madsen}, \citenamefont {Kvasnicka},\
  and\ \citenamefont {Luitz}}]{Wien2k}%
  \BibitemOpen
  \bibfield  {author} {\bibinfo {author} {\bibfnamefont {P.}~\bibnamefont
  {Blaha}}, \bibinfo {author} {\bibfnamefont {K.}~\bibnamefont {Schwarz}},
  \bibinfo {author} {\bibfnamefont {G.~K.~H.}\ \bibnamefont {Madsen}}, \bibinfo
  {author} {\bibfnamefont {D.}~\bibnamefont {Kvasnicka}}, \ and\ \bibinfo
  {author} {\bibfnamefont {J.}~\bibnamefont {Luitz}},\ }\href@noop {}
  {\bibfield  {journal} {\bibinfo  {journal} {WIEN2k, An Augmented Plane Wave
  Plus Local Orbitals Program for Calculating Crystal Properties ({K}arlheinz
  {S}chwarz, {T}echn. {U}niversit{\"a}t {W}ien, {A}ustria)}\ } (\bibinfo {year}
  {2001})}\BibitemShut {NoStop}%
\bibitem [{\citenamefont {Winter}\ \emph
  {et~al.}(2017{\natexlab{c}})\citenamefont {Winter}, \citenamefont {Riedl},\
  and\ \citenamefont {Valent\'{\i}}}]{Winter2017CT}%
  \BibitemOpen
  \bibfield  {author} {\bibinfo {author} {\bibfnamefont {S.~M.}\ \bibnamefont
  {Winter}}, \bibinfo {author} {\bibfnamefont {K.}~\bibnamefont {Riedl}}, \
  and\ \bibinfo {author} {\bibfnamefont {R.}~\bibnamefont {Valent\'{\i}}},\
  }\href@noop {} {\bibfield  {journal} {\bibinfo  {journal} {Phys. Rev. B}\
  }\textbf {\bibinfo {volume} {95}},\ \bibinfo {pages} {060404} (\bibinfo
  {year} {2017}{\natexlab{c}})}\BibitemShut {NoStop}%
\bibitem [{Note1()}]{Note1}%
  \BibitemOpen
  \bibinfo {note} {See Supplemental Material which contains structural
  parameters, density of states, and hopping parameters for various structures
  as well as Ref~\protect \rev@citealp {Wien2k, Bloechl1994, Perdew1996,
  Kresse1996, Hafner2008, Foyevtsova2013, Rau2014}}\BibitemShut {NoStop}%
\bibitem [{\citenamefont {Chaloupka}\ and\ \citenamefont
  {Khaliullin}(2015)}]{chaloupka2015hidden}%
  \BibitemOpen
  \bibfield  {author} {\bibinfo {author} {\bibfnamefont {J.}~\bibnamefont
  {Chaloupka}}\ and\ \bibinfo {author} {\bibfnamefont {G.}~\bibnamefont
  {Khaliullin}},\ }\href@noop {} {\bibfield  {journal} {\bibinfo  {journal}
  {Physical Review B}\ }\textbf {\bibinfo {volume} {92}},\ \bibinfo {pages}
  {024413} (\bibinfo {year} {2015})}\BibitemShut {NoStop}%
\bibitem [{\citenamefont {McKenzie}\ \emph {et~al.}(2014)\citenamefont
  {McKenzie}, \citenamefont {Bekker}, \citenamefont {Athokpam},\ and\
  \citenamefont {Ramesh}}]{mckenzie2014effect}%
  \BibitemOpen
  \bibfield  {author} {\bibinfo {author} {\bibfnamefont {R.~H.}\ \bibnamefont
  {McKenzie}}, \bibinfo {author} {\bibfnamefont {C.}~\bibnamefont {Bekker}},
  \bibinfo {author} {\bibfnamefont {B.}~\bibnamefont {Athokpam}}, \ and\
  \bibinfo {author} {\bibfnamefont {S.~G.}\ \bibnamefont {Ramesh}},\
  }\href@noop {} {\bibfield  {journal} {\bibinfo  {journal} {The Journal of
  chemical physics}\ }\textbf {\bibinfo {volume} {140}},\ \bibinfo {pages}
  {174508} (\bibinfo {year} {2014})}\BibitemShut {NoStop}%
\bibitem [{\citenamefont {Perrin}\ and\ \citenamefont
  {Nielson}(1997)}]{Charles1997}%
  \BibitemOpen
  \bibfield  {author} {\bibinfo {author} {\bibfnamefont {C.~L.}\ \bibnamefont
  {Perrin}}\ and\ \bibinfo {author} {\bibfnamefont {J.~B.}\ \bibnamefont
  {Nielson}},\ }\href@noop {} {\bibfield  {journal} {\bibinfo  {journal}
  {Annual Review of Physical Chemistry}\ }\textbf {\bibinfo {volume} {48}},\
  \bibinfo {pages} {511} (\bibinfo {year} {1997})}\BibitemShut {NoStop}%
\bibitem [{\citenamefont {Schi{\o}tt}\ \emph {et~al.}(1998)\citenamefont
  {Schi{\o}tt}, \citenamefont {Iversen}, \citenamefont {Madsen}, \citenamefont
  {Larsen},\ and\ \citenamefont {Bruice}}]{Schiott1998}%
  \BibitemOpen
  \bibfield  {author} {\bibinfo {author} {\bibfnamefont {B.}~\bibnamefont
  {Schi{\o}tt}}, \bibinfo {author} {\bibfnamefont {B.~B.}\ \bibnamefont
  {Iversen}}, \bibinfo {author} {\bibfnamefont {G.~K.~H.}\ \bibnamefont
  {Madsen}}, \bibinfo {author} {\bibfnamefont {F.~K.}\ \bibnamefont {Larsen}},
  \ and\ \bibinfo {author} {\bibfnamefont {T.~C.}\ \bibnamefont {Bruice}},\
  }\href@noop {} {\bibfield  {journal} {\bibinfo  {journal} {Proceedings of the
  National Academy of Sciences}\ }\textbf {\bibinfo {volume} {95}},\ \bibinfo
  {pages} {12799} (\bibinfo {year} {1998})}\BibitemShut {NoStop}%
\bibitem [{\citenamefont {Christensen}\ \emph {et~al.}(1977)\citenamefont
  {Christensen}, \citenamefont {Hansen},\ and\ \citenamefont
  {Lehmann}}]{christensen1977isotope}%
  \BibitemOpen
  \bibfield  {author} {\bibinfo {author} {\bibfnamefont {A.~N.}\ \bibnamefont
  {Christensen}}, \bibinfo {author} {\bibfnamefont {P.}~\bibnamefont {Hansen}},
  \ and\ \bibinfo {author} {\bibfnamefont {M.}~\bibnamefont {Lehmann}},\
  }\href@noop {} {\bibfield  {journal} {\bibinfo  {journal} {Journal of Solid
  State Chemistry}\ }\textbf {\bibinfo {volume} {21}},\ \bibinfo {pages} {325}
  (\bibinfo {year} {1977})}\BibitemShut {NoStop}%
\bibitem [{\citenamefont {Matsuo}\ \emph {et~al.}(2000)\citenamefont {Matsuo},
  \citenamefont {Inaba}, \citenamefont {Yamamuro},\ and\ \citenamefont
  {Onoda-Yamamuro}}]{matsuo2000proton}%
  \BibitemOpen
  \bibfield  {author} {\bibinfo {author} {\bibfnamefont {T.}~\bibnamefont
  {Matsuo}}, \bibinfo {author} {\bibfnamefont {A.}~\bibnamefont {Inaba}},
  \bibinfo {author} {\bibfnamefont {O.}~\bibnamefont {Yamamuro}}, \ and\
  \bibinfo {author} {\bibfnamefont {N.}~\bibnamefont {Onoda-Yamamuro}},\
  }\href@noop {} {\bibfield  {journal} {\bibinfo  {journal} {Journal of
  Physics: Condensed Matter}\ }\textbf {\bibinfo {volume} {12}},\ \bibinfo
  {pages} {8595} (\bibinfo {year} {2000})}\BibitemShut {NoStop}%
\bibitem [{\citenamefont {Matsuo}\ \emph {et~al.}(2006)\citenamefont {Matsuo},
  \citenamefont {Maekawa}, \citenamefont {Inaba}, \citenamefont {Yamamuro},
  \citenamefont {Ohama}, \citenamefont {Ichikawa},\ and\ \citenamefont
  {Tsuchida}}]{matsuo2006isotope}%
  \BibitemOpen
  \bibfield  {author} {\bibinfo {author} {\bibfnamefont {T.}~\bibnamefont
  {Matsuo}}, \bibinfo {author} {\bibfnamefont {T.}~\bibnamefont {Maekawa}},
  \bibinfo {author} {\bibfnamefont {A.}~\bibnamefont {Inaba}}, \bibinfo
  {author} {\bibfnamefont {O.}~\bibnamefont {Yamamuro}}, \bibinfo {author}
  {\bibfnamefont {M.}~\bibnamefont {Ohama}}, \bibinfo {author} {\bibfnamefont
  {M.}~\bibnamefont {Ichikawa}}, \ and\ \bibinfo {author} {\bibfnamefont
  {T.}~\bibnamefont {Tsuchida}},\ }\href@noop {} {\bibfield  {journal}
  {\bibinfo  {journal} {Journal of molecular structure}\ }\textbf {\bibinfo
  {volume} {790}},\ \bibinfo {pages} {129} (\bibinfo {year}
  {2006})}\BibitemShut {NoStop}%
\bibitem [{\citenamefont {Dolin}\ \emph {et~al.}(2007)\citenamefont {Dolin},
  \citenamefont {Flyagina}, \citenamefont {Tremasova}, \citenamefont
  {Mikhailova}, \citenamefont {Gavrilyuk},\ and\ \citenamefont
  {Levin}}]{dolin2007study}%
  \BibitemOpen
  \bibfield  {author} {\bibinfo {author} {\bibfnamefont {S.}~\bibnamefont
  {Dolin}}, \bibinfo {author} {\bibfnamefont {I.}~\bibnamefont {Flyagina}},
  \bibinfo {author} {\bibfnamefont {M.}~\bibnamefont {Tremasova}}, \bibinfo
  {author} {\bibfnamefont {T.~Y.}\ \bibnamefont {Mikhailova}}, \bibinfo
  {author} {\bibfnamefont {A.}~\bibnamefont {Gavrilyuk}}, \ and\ \bibinfo
  {author} {\bibfnamefont {A.}~\bibnamefont {Levin}},\ }\href@noop {}
  {\bibfield  {journal} {\bibinfo  {journal} {International Journal of Quantum
  Chemistry}\ }\textbf {\bibinfo {volume} {107}},\ \bibinfo {pages} {2409}
  (\bibinfo {year} {2007})}\BibitemShut {NoStop}%
\bibitem [{\citenamefont {Ueda}\ \emph {et~al.}(2014)\citenamefont {Ueda},
  \citenamefont {Yamada}, \citenamefont {Isono}, \citenamefont {Kamo},
  \citenamefont {Nakao}, \citenamefont {Kumai}, \citenamefont {Nakao},
  \citenamefont {Murakami}, \citenamefont {Yamamoto}, \citenamefont {Nishio}
  \emph {et~al.}}]{ueda2014hydrogen}%
  \BibitemOpen
  \bibfield  {author} {\bibinfo {author} {\bibfnamefont {A.}~\bibnamefont
  {Ueda}}, \bibinfo {author} {\bibfnamefont {S.}~\bibnamefont {Yamada}},
  \bibinfo {author} {\bibfnamefont {T.}~\bibnamefont {Isono}}, \bibinfo
  {author} {\bibfnamefont {H.}~\bibnamefont {Kamo}}, \bibinfo {author}
  {\bibfnamefont {A.}~\bibnamefont {Nakao}}, \bibinfo {author} {\bibfnamefont
  {R.}~\bibnamefont {Kumai}}, \bibinfo {author} {\bibfnamefont
  {H.}~\bibnamefont {Nakao}}, \bibinfo {author} {\bibfnamefont
  {Y.}~\bibnamefont {Murakami}}, \bibinfo {author} {\bibfnamefont
  {K.}~\bibnamefont {Yamamoto}}, \bibinfo {author} {\bibfnamefont
  {Y.}~\bibnamefont {Nishio}},  \emph {et~al.},\ }\href@noop {} {\bibfield
  {journal} {\bibinfo  {journal} {Journal of the American Chemical Society}\
  }\textbf {\bibinfo {volume} {136}},\ \bibinfo {pages} {12184} (\bibinfo
  {year} {2014})}\BibitemShut {NoStop}%
\bibitem [{\citenamefont {Yamamoto}\ \emph {et~al.}(2016)\citenamefont
  {Yamamoto}, \citenamefont {Kanematsu}, \citenamefont {Nagashima},
  \citenamefont {Ueda}, \citenamefont {Mori},\ and\ \citenamefont
  {Tachikawa}}]{yamamoto2016theoretical}%
  \BibitemOpen
  \bibfield  {author} {\bibinfo {author} {\bibfnamefont {K.}~\bibnamefont
  {Yamamoto}}, \bibinfo {author} {\bibfnamefont {Y.}~\bibnamefont {Kanematsu}},
  \bibinfo {author} {\bibfnamefont {U.}~\bibnamefont {Nagashima}}, \bibinfo
  {author} {\bibfnamefont {A.}~\bibnamefont {Ueda}}, \bibinfo {author}
  {\bibfnamefont {H.}~\bibnamefont {Mori}}, \ and\ \bibinfo {author}
  {\bibfnamefont {M.}~\bibnamefont {Tachikawa}},\ }\href@noop {} {\bibfield
  {journal} {\bibinfo  {journal} {Physical Chemistry Chemical Physics}\
  }\textbf {\bibinfo {volume} {18}},\ \bibinfo {pages} {29673} (\bibinfo {year}
  {2016})}\BibitemShut {NoStop}%
\bibitem [{\citenamefont {Kimchi}\ \emph
  {et~al.}(2018{\natexlab{b}})\citenamefont {Kimchi}, \citenamefont {Nahum},\
  and\ \citenamefont {Senthil}}]{Kimchi2018Yb}%
  \BibitemOpen
  \bibfield  {author} {\bibinfo {author} {\bibfnamefont {I.}~\bibnamefont
  {Kimchi}}, \bibinfo {author} {\bibfnamefont {A.}~\bibnamefont {Nahum}}, \
  and\ \bibinfo {author} {\bibfnamefont {T.}~\bibnamefont {Senthil}},\ }\href
  {\doibase 10.1103/PhysRevX.8.031028} {\bibfield  {journal} {\bibinfo
  {journal} {Phys. Rev. X}\ }\textbf {\bibinfo {volume} {8}},\ \bibinfo {pages}
  {031028} (\bibinfo {year} {2018}{\natexlab{b}})}\BibitemShut {NoStop}%
\bibitem [{\citenamefont {Willans}\ \emph {et~al.}(2010)\citenamefont
  {Willans}, \citenamefont {Chalker},\ and\ \citenamefont
  {Moessner}}]{Willans2010}%
  \BibitemOpen
  \bibfield  {author} {\bibinfo {author} {\bibfnamefont {A.~J.}\ \bibnamefont
  {Willans}}, \bibinfo {author} {\bibfnamefont {J.~T.}\ \bibnamefont
  {Chalker}}, \ and\ \bibinfo {author} {\bibfnamefont {R.}~\bibnamefont
  {Moessner}},\ }\href {\doibase 10.1103/PhysRevLett.104.237203} {\bibfield
  {journal} {\bibinfo  {journal} {Phys. Rev. Lett.}\ }\textbf {\bibinfo
  {volume} {104}},\ \bibinfo {pages} {237203} (\bibinfo {year}
  {2010})}\BibitemShut {NoStop}%
\bibitem [{\citenamefont {Paik}\ \emph {et~al.}(2002)\citenamefont {Paik},
  \citenamefont {Grey}, \citenamefont {Johnson}, \citenamefont {Kim},\ and\
  \citenamefont {Thackeray}}]{paik2002lithium}%
  \BibitemOpen
  \bibfield  {author} {\bibinfo {author} {\bibfnamefont {Y.}~\bibnamefont
  {Paik}}, \bibinfo {author} {\bibfnamefont {C.~P.}\ \bibnamefont {Grey}},
  \bibinfo {author} {\bibfnamefont {C.~S.}\ \bibnamefont {Johnson}}, \bibinfo
  {author} {\bibfnamefont {J.-S.}\ \bibnamefont {Kim}}, \ and\ \bibinfo
  {author} {\bibfnamefont {M.~M.}\ \bibnamefont {Thackeray}},\ }\href@noop {}
  {\bibfield  {journal} {\bibinfo  {journal} {Chemistry of materials}\ }\textbf
  {\bibinfo {volume} {14}},\ \bibinfo {pages} {5109} (\bibinfo {year}
  {2002})}\BibitemShut {NoStop}%
\bibitem [{\citenamefont {Tang}\ \emph {et~al.}(2000)\citenamefont {Tang},
  \citenamefont {Kanoh}, \citenamefont {Yang},\ and\ \citenamefont
  {Ooi}}]{tang2000preparation}%
  \BibitemOpen
  \bibfield  {author} {\bibinfo {author} {\bibfnamefont {W.}~\bibnamefont
  {Tang}}, \bibinfo {author} {\bibfnamefont {H.}~\bibnamefont {Kanoh}},
  \bibinfo {author} {\bibfnamefont {X.}~\bibnamefont {Yang}}, \ and\ \bibinfo
  {author} {\bibfnamefont {K.}~\bibnamefont {Ooi}},\ }\href@noop {} {\bibfield
  {journal} {\bibinfo  {journal} {Chemistry of materials}\ }\textbf {\bibinfo
  {volume} {12}},\ \bibinfo {pages} {3271} (\bibinfo {year}
  {2000})}\BibitemShut {NoStop}%
\bibitem [{\citenamefont {Weber}\ \emph {et~al.}(2017)\citenamefont {Weber},
  \citenamefont {Schoop}, \citenamefont {Wurmbrand}, \citenamefont {Nuss},
  \citenamefont {Seibel}, \citenamefont {Tafti}, \citenamefont {Ji},
  \citenamefont {Cava}, \citenamefont {Dinnebier},\ and\ \citenamefont
  {Lotsch}}]{weber2017trivalent}%
  \BibitemOpen
  \bibfield  {author} {\bibinfo {author} {\bibfnamefont {D.}~\bibnamefont
  {Weber}}, \bibinfo {author} {\bibfnamefont {L.~M.}\ \bibnamefont {Schoop}},
  \bibinfo {author} {\bibfnamefont {D.}~\bibnamefont {Wurmbrand}}, \bibinfo
  {author} {\bibfnamefont {J.}~\bibnamefont {Nuss}}, \bibinfo {author}
  {\bibfnamefont {E.~M.}\ \bibnamefont {Seibel}}, \bibinfo {author}
  {\bibfnamefont {F.~F.}\ \bibnamefont {Tafti}}, \bibinfo {author}
  {\bibfnamefont {H.}~\bibnamefont {Ji}}, \bibinfo {author} {\bibfnamefont
  {R.~J.}\ \bibnamefont {Cava}}, \bibinfo {author} {\bibfnamefont {R.~E.}\
  \bibnamefont {Dinnebier}}, \ and\ \bibinfo {author} {\bibfnamefont {B.~V.}\
  \bibnamefont {Lotsch}},\ }\href@noop {} {\bibfield  {journal} {\bibinfo
  {journal} {Chemistry of Materials}\ }\textbf {\bibinfo {volume} {29}},\
  \bibinfo {pages} {8338} (\bibinfo {year} {2017})}\BibitemShut {NoStop}%
\bibitem [{\citenamefont {Bl\"ochl}(1994)}]{Bloechl1994}%
  \BibitemOpen
  \bibfield  {author} {\bibinfo {author} {\bibfnamefont {P.~E.}\ \bibnamefont
  {Bl\"ochl}},\ }\href@noop {} {\bibfield  {journal} {\bibinfo  {journal}
  {Phys. Rev. B}\ }\textbf {\bibinfo {volume} {50}},\ \bibinfo {pages} {17953}
  (\bibinfo {year} {1994})}\BibitemShut {NoStop}%
\bibitem [{\citenamefont {Perdew}\ \emph {et~al.}(1996)\citenamefont {Perdew},
  \citenamefont {Burke},\ and\ \citenamefont {Ernzerhof}}]{Perdew1996}%
  \BibitemOpen
  \bibfield  {author} {\bibinfo {author} {\bibfnamefont {J.~P.}\ \bibnamefont
  {Perdew}}, \bibinfo {author} {\bibfnamefont {K.}~\bibnamefont {Burke}}, \
  and\ \bibinfo {author} {\bibfnamefont {M.}~\bibnamefont {Ernzerhof}},\
  }\href@noop {} {\bibfield  {journal} {\bibinfo  {journal} {Phys. Rev. Lett.}\
  }\textbf {\bibinfo {volume} {77}},\ \bibinfo {pages} {3865} (\bibinfo {year}
  {1996})}\BibitemShut {NoStop}%
\end{thebibliography}%

\end{document}